\documentclass[
 reprint,
 amsmath,amssymb,
 aps,
]{revtex4-1}

\usepackage{graphicx}% Include figure files
\usepackage{dcolumn}% Align table columns on decimal point
\usepackage{bm}% bold math
\newcommand{\Lagr}{\mathop{\mathcal{L}}}

\begin{document}

\rightline{SESAPS\,-2015\,-\,F1\,-4}

\title{Exploration of a High Luminosity 100 TeV Proton Antiproton Collider}
\author{S.\,J.\,Oliveros, D.\,J.\,Summers, L.\,M.\,Cremaldi, and J.\,G.\,Acosta}
\thanks{solivero@go.olemiss.edu}
\affiliation{University of Mississippi\,-\,Oxford, University, MS 38677 \ USA}
\author{D. V. Neuffer}
\affiliation{Fermilab, Batavia, IL 60510 USA}

\date{\today}

\begin{abstract}
New physics is being explored with the Large Hadron Collider at CERN and with Intensity Frontier programs at Fermilab and KEK. The energy scale for new physics is known to be in the multi-TeV range, signaling the need for a future collider which well surpasses this energy scale. We explore  a 10$^{\,34}$ cm$^{-2}$s$^{-1}$ luminosity, 100 TeV $p\bar{p}$ collider with 7$\times$ the energy of the LHC but only 2$\times$ as much NbTi superconductor, motivating the choice of 4.5\,T single bore dipoles. The cross section for many high mass states is 10 times higher in $p\bar{p}$ than $pp$ collisions. Antiquarks for production can come directly from an antiproton rather than indirectly from gluon splitting. The higher cross sections reduce the synchrotron radiation in superconducting magnets and the number of events per beam crossing, because lower beam currents can produce the same rare event rates. Events are more centrally produced, allowing a more compact detector with less space between quadrupole triplets and a smaller $\beta^{*}$ for higher luminosity. A Fermilab-like $\bar p$ source would disperse the beam into 12 momentum channels to capture more antiprotons. Because stochastic cooling time scales as the number of particles, 12 cooling ring sets would be used. Each set would include  phase rotation to lower momentum spreads, equalize all momentum channels, and stochastically cool. One electron cooling ring would follow the stochastic cooling rings. Finally antiprotons would be recycled during runs without leaving the collider ring by joining them to new bunches with synchrotron damping. 
\end{abstract}

\maketitle

\section{Introduction}
With the recent discovery of the Higgs boson the standard model of particle physics is complete, but exploration will continue to search for beyond the standard model (BSM) physics that many agree must exist. There is still no explanation for dark matter\,\cite{Zwicky}. The CMS and ATLAS experiments at the LHC (Large Hadron Collider) observed a Higgs boson with mass of about 125 GeV, using a data sample with collision energies of 7 and 8 TeV\,\cite{Higgs}. The LHC started taking data at 13 TeV in 2015, and it will continue increasing the number of collisions. The LHC design luminosity of 10$^{34}$ cm$^{-2}$s$^{-1}$ was reached in June 2016. 

The second most powerful hadron collider was the Tevatron at Fermilab. This collided protons and antiprotons with a center of mass energy of 1.96 TeV. During its 28 years of operation (closing in 2011) one of its greatest discoveries was the top quark in 1995\,\cite{top}. New Fermilab particle research projects are under development. Among them are new high intensity neutrino and muon decay experiments.

Many agree that hadron colliders beyond 14 TeV are necessary to fully explore new BSM physics. For that reason, a 100 TeV proton-proton collider in a 100 km ring with 16\,T Nb$_{\,3}$Sn dipoles is being considered for CERN. Synchrotron radiation in the dipoles and vacuum are concerns. The design is challenging. Might a simpler hadron collider be possible? This is the motivation to consider a 100 TeV  proton-antiproton ($p\bar{p}$) collider with a luminosity of 10$^{\,34}$ cm$^{-2}$s$^{-1}$ based on 4.5\,T NbTi superferric dipoles\,\cite{McIntyre2015} in a 270\,km tunnel. Much of the technology for this collider is available and was demonstrated at the Tevatron ($\Lagr$ = 4 x 10$^{32}$ cm$^{-2}$s$^{-1}$). The most important factor in the study of a 100 TeV $p\bar{p}$ machine is to find a way to cool and recycle more antiprotons to achieve higher luminosity.

\section{Proton antiproton collider remarks}
Proton antiproton colliders have been used at CERN\,\cite{CERN}, Fermilab\,\cite{Fermilab}, and GSI Darmstadt\,\cite{GSI}. Also, the SSC (Superconducting Super Collider) Central Design Group\,\cite{Barish:1986kj} presented studies which examined the option of a $p\bar{p}$ collider ring for the SSC Super Collider in Texas. 

\subsection{High mass cross sections in $pp$ and $p\bar{p}$ collisions}
\begin{figure}[t]
	\centering
	\includegraphics*[width=85mm]{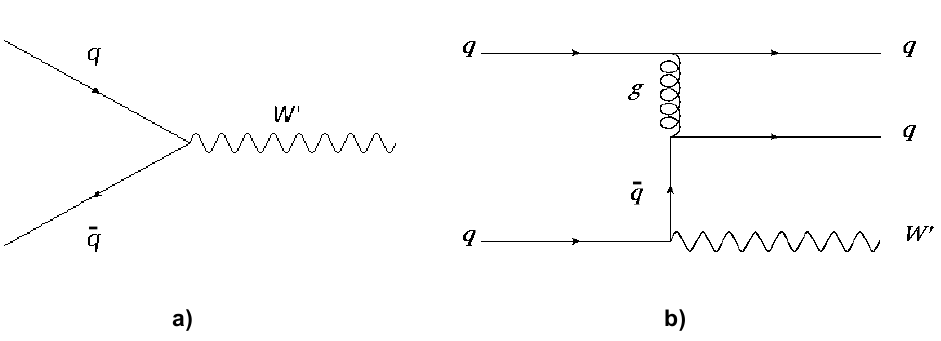}
	\caption{Feynman diagrams for W$'$ production in (a) $q\bar{q}$ collision (s channel), and (b) $qq$ collision (t channel). The two final state quarks cross in the u channel, which is not shown.}
	\label{fig:Fig1}
\end{figure}

Physics beyond the standard model is of great interest. New particles are predicted by unified theories. As an example, one of them is the hypothetical boson $W'$, which is a massive version of the standard model $W$ boson. This predicted boson is considered to have a mass around the TeV scale, and to be produced through $q\bar{q}$ collisions. Fig.\,\ref{fig:Fig1} shows the Feynman diagram for $W'$ production from $q\bar{q}$ and $qq$ annihilation. There, it can be seen that antiquarks for $W'$ production can come directly from an antiproton rather than indirectly from gluon splitting in proton-proton collisions.

\begin{figure}[b]
	\centering
	\includegraphics*[width=85mm]{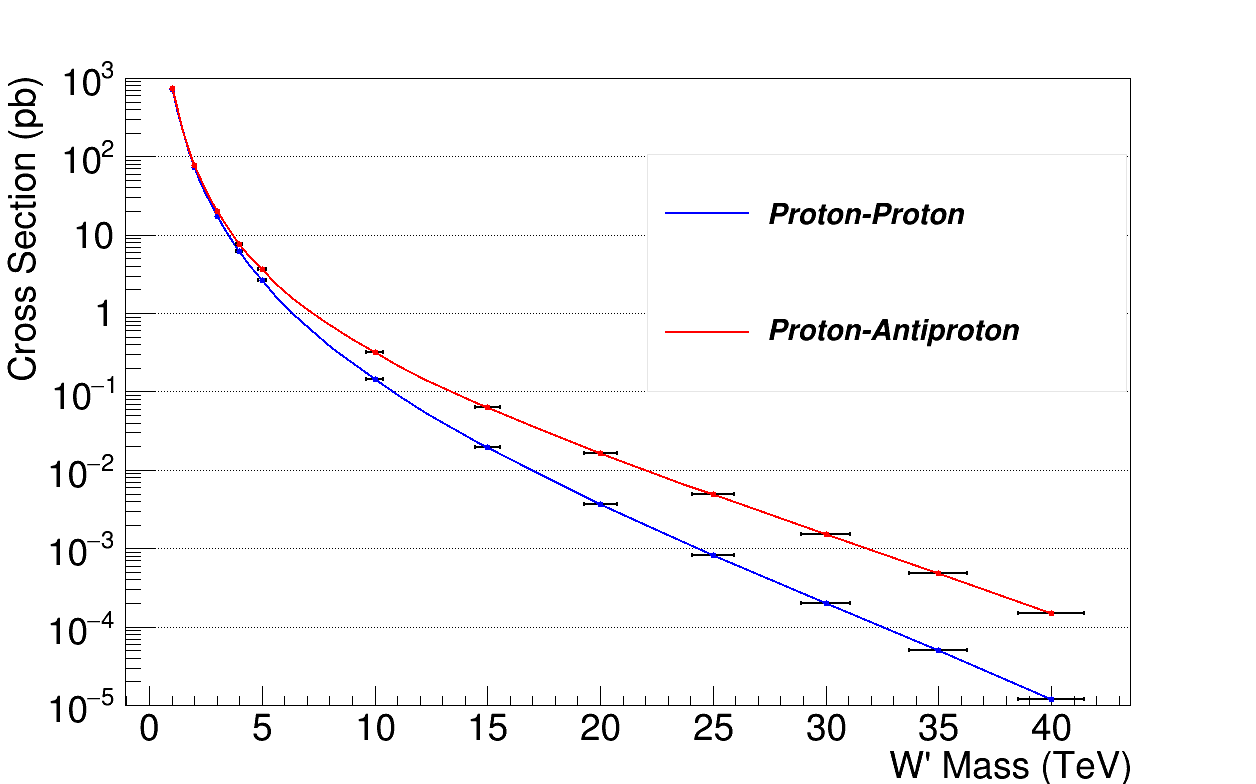}
	\caption{$W'$ Boson production cross section as a function of the mass using $pp$ and $p\bar{p}$ collisions with a E$_{cm}$ = 100 Tev.}
	\label{fig:Fig2}
\end{figure}

Some high mass cross sections for $p\bar{p}$ collisions are greater than for $pp$ collisions~\cite{Summers}. This can be corroborated by calculating the cross section for $W'$ production for different masses. Using an event generator, Madgraph\,\cite{Alwall:2014hca}, the $W'$ cross section is obtained for different $W'$ masses using proton-proton and proton-antiproton collisions at a center of mass energy of 100 TeV. The results are shown in the Fig.\,\ref{fig:Fig2}. As the mass increases the $W'$ cross section obtained with $p\bar{p}$ collisions is greater compared to $pp$ collisions, becoming around 10 times larger at higher masses.

A great advantage of having higher cross sections is that this allows the collider to be run at lower luminosities generating  less detector radiation damage and less detector pileup. Pileup decreases as antiprotons are distributed into more bunches.  
There is a limit to this as the number of protons per bunch must stay constant to keep the luminosity constant. Synchrotron radiation rises with the number of protons. In equation 1, $\gamma$ is the relativistic factor, N is the number of particles per bunch, f is the collision frequency which increases linearly with the number of bunches, $\mathrm{\epsilon_N}$ is the normalized transverse emittance, and $\beta^*$ is the twiss function\,\cite{Courant} at the interaction point (IP).
\begin{equation}
%\begin{align}
\mathrm{{\Lagr} = {{\gamma \, N_{\bar{p}} N_p \, f} \over {4 \pi \, \epsilon_N \, \beta^*}}}
\label{luminosity}
%\end{aligned}
\end{equation}

\subsection{Synchrotron radiation in high energy colliders}
It is important to note that the higher cross sections for high $p_T$ particle production at $p\bar{p}$ colliders allow the collider to run at lower beam currents and luminosities, reducing synchrotron radiation in the collider's superconducting magnets and vacuum system. To estimate how much power is emitted due to Synchrotron Radiation (SR), the equation\,\cite{Barletta-uspas},
\begin{equation}
P = \frac{c}{6 \pi \epsilon_0} \, q^{\,2} \, \frac{(\beta \, \gamma)^4}{\rho \, R}
\label{power}
\end{equation}
\noindent
gives the power emitted per particle, where $\epsilon_o = 8.85 \times 10^{-12}$ farads/meter, $q = 1.6 \times 10^{-19}$ coulombs, $\rho$ is bend radius, and $R$ is ring radius. For example, in the LHC (E~=~7\,TeV, $\gamma$\,=\,E/m\,=\,7460, $\rho=2800$\,m, R\,=\,4300\,m) the SR power for a proton is $P = 1.2\times10^{-14}$ kW. Thus, for a proton beam, which has 2808 bunches and $1.15\times10^{11}$ protons per bunch, the total SR radiated is 3.9 kW. Now, if the particle energy is increased, the SR power grows as $\gamma^{4}/\rho R$, presenting a problem to deal with in designing a future high energy (100 TeV) collider\,\cite{Barletta}. Table\,\ref{SR} shows the parameters and the SR power calculations for a 100 km $pp$ collider\,\cite{Zimmermann:2015jpa}, and for the 270 km $p\bar{p}$ collider proposed in this work. 

\begin{table*}[t]
\renewcommand{\arraystretch}{1.15}
\tabcolsep=1.4mm
	\centering
	\caption{Synchrotron Radiation (SR) for a 100 km $pp$ and 270 km $p\bar{p}$ circumference colliders. $\gamma$ = E/m = 53,300.}
	\label{SR}
\begin{ruledtabular}
   		\begin{tabular}{ccccccccccccc}
			Collider & R & E$_{beam}$ & B & $\rho$ & Packing & SR/proton & Bunches & Particles & SR/pipe & Beam & SR/meter \\ 
			& m & TeV & Tesla & m & Fraction & kW & per beam & per bunch & kW & Pipes & W/m \\\hline
			$pp$ & 15,915 & 50 & 16 & 10,410 & 0.66 & $2.25 \times 10^{-12}$ & 10,600 & $10 \times 10^{10}$ & 2380 & 2 & 29 \\ 
			$p\bar{p}$ & 42,970 & 50 & 4.5 & 37,040 & 0.86 & $2.33 \times 10^{-13}$ & 10,800 & $20/0.32 \times 10^{10}$ & 511 & 1 & 2.2 
			\\ 
		\end{tabular}
		\end{ruledtabular}		
	%	}
\end{table*}
It is important to highlight that the SR Power per meter is 13 times lower for the 100 TeV $p\bar{p}$ compared with the $pp$ collider, because higher production cross sections of high $p_T$ particles allow lower beam currents and because the tunnel circumference is larger. Furthermore, a $p\bar{p}$ collider only requires one ring instead of the two needed for a $pp$ collider. The same magnets are shared by both beams because of their opposite charge, reducing costs. All reasons above are advantages of a high energy proton-antiproton collider. 

\section{Antiprotons capture}
A requirement  in designing the 100 TeV  $p\bar{p}$ collider will be achieving a luminosity of 10$^{\,34}\,\mathrm{cm^{-2}s^{-1}}$. To increase the luminosity in $\textit{p}\bar{p}$ colliders, the number of antiprotons is a crucial factor. Fermilab had a powerful antiproton source, which cooled about 25 x 10$^{10}$ $\bar{p}$/hour with a momentum of 8.9 $\pm\,2\%$ GeV/c in the Debuncher and Accumulator rings. A large fraction of antiprotons were rejected because of the momentum acceptance of only $\pm\,2\,\%$. As a starting point, taking as reference the Tevatron collider, the gain in luminosity for the 100 TeV $p\bar{p}$ collider, for which the beam energy is 50 TeV, the ring circumference  270 km, and with $\beta^{*}$ = 14 cm (half of the Tevatron), can be scaled as,
\begin{equation}
\begin{aligned}
\mathrm{{\Lagr}_{scaled} = E_{increased} \times f_{decreased} \times \beta^{*} _{factor} \times {\Lagr}_{current}} \\
\mathrm{= \frac{50\,TeV}{0.98\,TeV} \times \frac{6.28\,km}{270\,km} \times 2 \times (3.4\times 10^{32}\,cm^{-2}\,s^{-1})} \\
\mathrm{= 8.1 \times 10^{32}\,cm^{-2}\,s^{-1}}
\end{aligned}
\end{equation}
Thus, with 12x more bunches a luminosity of 10$^{34}$ can be  achieved. The antiproton burn rate for a 100 TeV $p\bar{p}$ collider, with total cross section $\sigma$= 153 mbarn \cite{Barletta:2013ooa} and ${\Lagr} = \mathrm{10^{\,34}\,cm^{-2}s^{-1}}$  is 
\begin{equation}
\mathrm{\bar{p}_{\,burn\,rate} = \sigma \cdot {\Lagr} = 551 \times 10^{10}  \,  \bar{p} /hr} 
\end{equation}
The Fermilab Debuncher cooled a peak of 45 x 10$^{10}$ $\bar{p}$/hr, thus the number of antiprotons needed are 12 times more,
at this peak rate. As previously mentioned, in the Fermilab antiproton source a large fraction of antiprotons were rejected because of the momentum acceptance. We then focus on collecting more of these antiprotons, specifically around 11 GeV/c $\pm\,24\,\%$. Providing a second Accumulator ring might improve the Accumulator ring stacking rate \cite{Lebedev}. At Fermilab, the Accumulator could not keep up with the Debuncher.

\subsection{Increase in antiproton momentum acceptance}
\begin{figure}[!htb]
	\centering
		\includegraphics*[width=85mm]{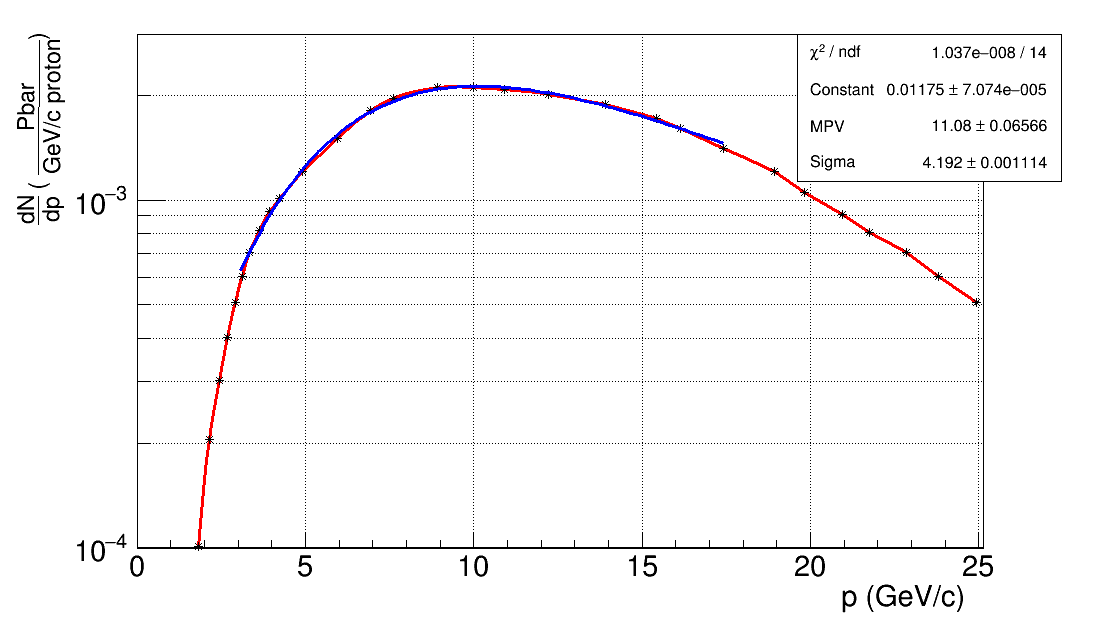}
		\caption{Momentum distribution of the antiprotons produced by a 120 GeV proton beam hitting a tungsten target \cite{Hojvat:1982xj}.}
		\label{fig:Fig3}
\end{figure}

To collect more antiprotons, a Fermilab-like target station would be used. Antiprotons were created by hitting an Inconel (a low expansion nickel-iron alloy) target with a spot size of $\sigma$ = 0.1 mm of 120 GeV protons. Fig.\,\ref{fig:Fig3} shows the momentum distribution for antiprotons created by a 120 GeV proton beam hitting a tungsten target within a production angle of 60 mrad. This plot was reproduced taking as reference  Figure 4 of the paper ``Calculation of anti-Proton Yields for the Fermilab anti-Proton Source''\,\cite{Hojvat:1982xj}. Inconel and tungsten should give similar distributions. The plot follows a Landau distribution function (blue trace) in the important momentum range of 5-18 GeV/c. Thus, the goal is to collect the antiprotons within an approximate momentum range of $p = 11.0\,\,$ GeV/c $\pm24 \%$ or $p = (11.0\,\pm\,2.6)\,\,$GeV/c, taking  11.0 GeV/c as the central momentum.

\begin{figure}[!htb]
	\centering
		\includegraphics*[width=85mm]{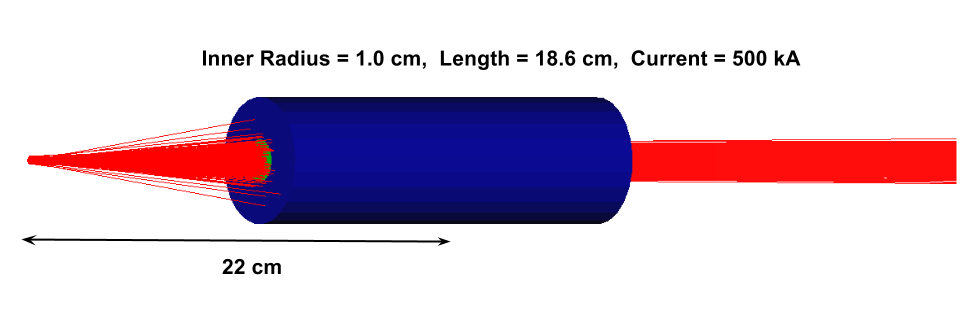}\\
		\includegraphics*[width=85mm]{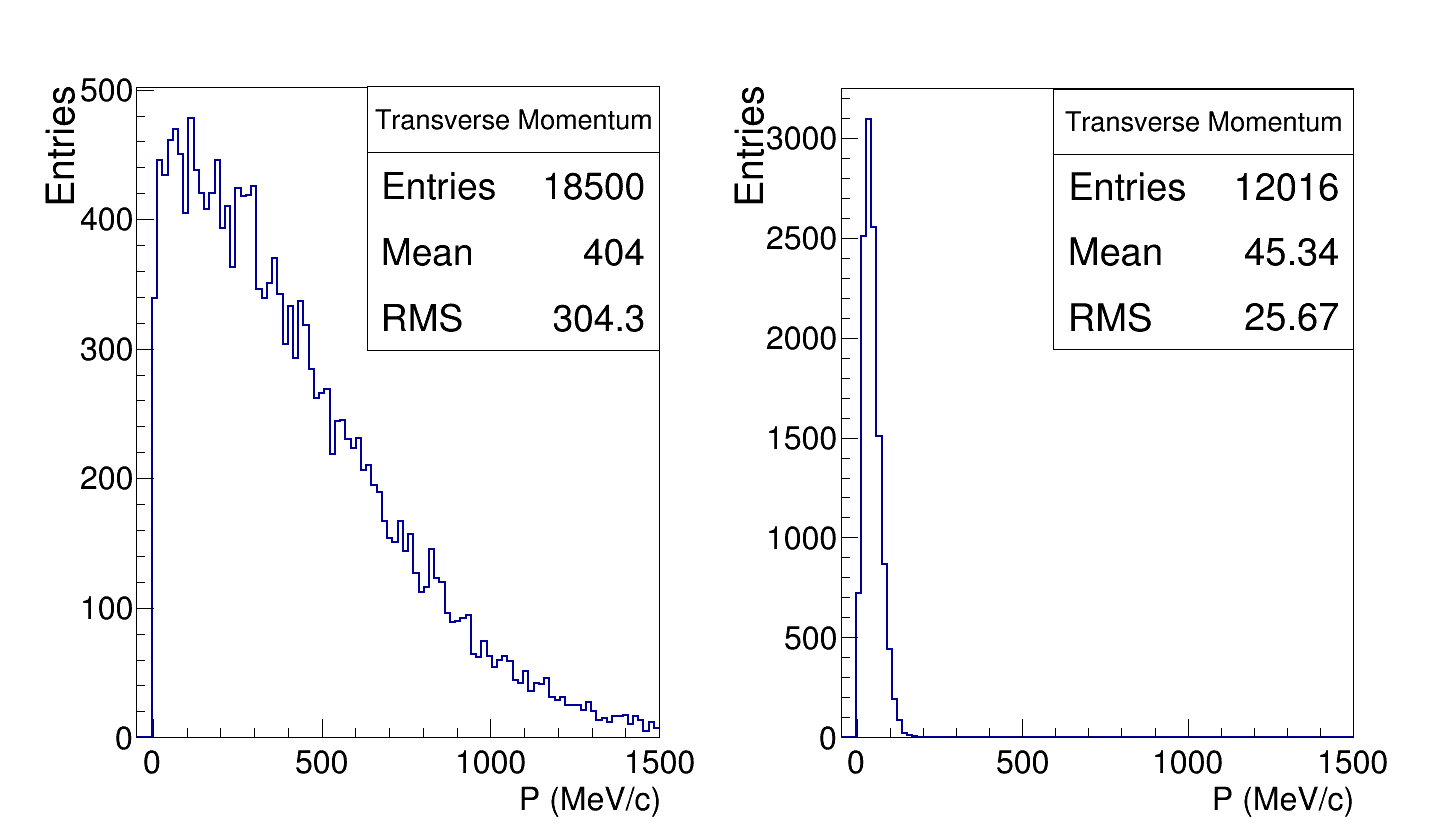}
		\caption{$11\,GeV/c \,\pm\,24\,\%$ antiproton beam simulation crossing a lithium lens in G4beamline. The plots show the transverse momentum distribution of the beam before entering (left plot) and in the output of the lithium lens (right plot)}
		\label{fig:Fig4}
\end{figure}

In the Tevatron target station, the particles emerged from the target and were focused by a lithium lens. This was a solid cylinder of radius 1 cm and length l5 cm, in which an high 500 kA axial current produced a strong radial gradient of 1,000 T/m. The effective focal length was 20 cm. To increase the momentum acceptance  of the antiprotons, a system of multiple lenses or a longer single lens might be necessary\,\cite{Childress}. To go from 8.9 to 11.0 GeV/c the lithium lens is lengthened from 15 to 18.6 cm. The new focal length for the central momentum of $11.0\,\,$ GeV/c is 20 cm. To measure the effectiveness of the new lens, a G4beamline\,\cite{G4bl} simulation was performed by creating a beam with a Landau momentum distribution (11 GeV/c,  $\sigma$ = 4 GeV/c, see Fig.\,\ref{fig:Fig3}), a gaussian angular distribution $\theta$\,(0,\,45\,mrad), and a uniform azimuthal distribution $\phi$ = (0, 2$\pi$). The purpose of the lithium lens is to reduce the transverse momentum, $p_{t}=\sqrt{p_x^{2}+p_y^{2}}$, of the beam. Fig.\,\ref{fig:Fig4} shows a simulated  11.0\ GeV/c $\pm24\%$ beam  coming out of a target and passing through a lithium lens. The transverse momentum of the beam before entering the lens, as well as the distribution after the lens is shown in the same figure. The mean transverse momentum value, $p_t$ is reduced 89$\%$, compared with a 91$\%$  reduction with the $p$ = 8.9 GeV/c $\pm\,2\%$ beam. The 18.6 cm lithium lens does a good job of reducing the transverse momentum of the beam with a large momentum spread.

In the lithium lens, the antiprotons experience scattering and absorption. The absorption causes loss and the scattering causes emittance growth. The deflection angle due to the Coulomb scattering\,\cite{Bichsel:2004ej} is given by,
\begin{equation} \label{eq:scattering}
\theta_{0}=\frac{13.6\,  {\rm MeV}}{\beta \, c\, p}z\sqrt{x/X_{0}}[1+0.038\,{\ln}(x/X_{0})]
\end{equation}
where $\beta$ = p/E =0.996, $p = 11\,GeV/c$, z = 1 (charge number) for the incident antiprotons, $x$ is the Li lens length (18.6 cm), and $X_{0}$ is the radiation length of the Li (15.5 cm). Substituting these values the scattering angle is 0.41 mrad. $\beta \, \gamma$ = p/m = 11.0/0.938 = 11.7. Thus, the normalized emittance growth \cite{McGinnis} $\mathrm{\Delta\epsilon_N  = \beta \, \gamma \, R_{Lens} \, \theta_{0}}$ is 48 $\mu$m.  The growth is small when added in quadrature to the initial normalized transverse beam emittance of 330~$\mu$m.

\subsection{Antiproton Beam Separation and Transport}
\begin{figure*}	
    \includegraphics[width=15cm]{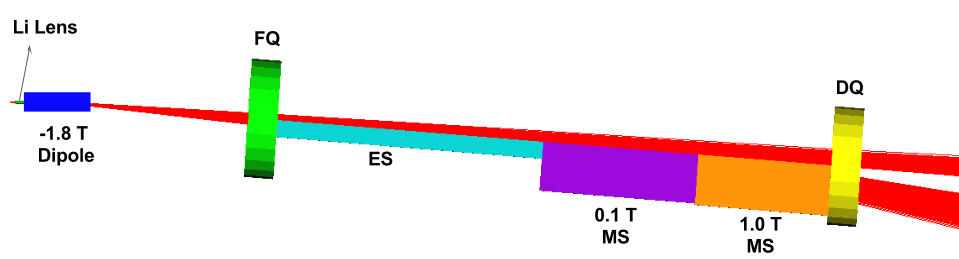}
	\caption{Configuration to divide the beam into two parts. An initial beam with momentum acceptance $p = 11.0\,\,$ GeV/c $\pm \,24 \%$ is collected by the Li lens and dispersed by a magnetic dipole to be then divided by a electrostatic septa ES and two magnetic dipoles MS.}
	\label{fig:Fig5}
\end{figure*}

\begin{figure*}	
	\includegraphics[width=15cm]{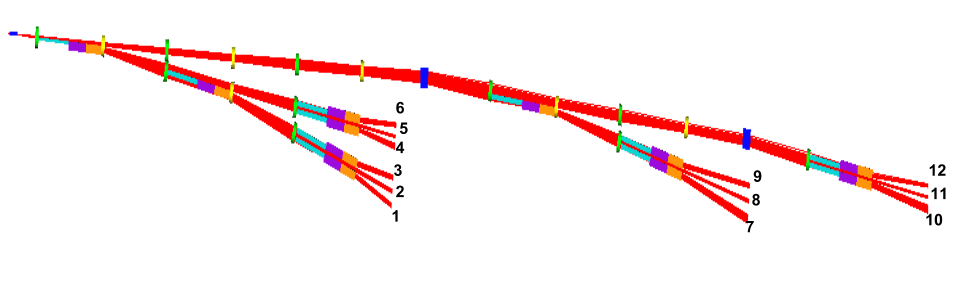}
	\centering
	\caption{An initial beam with momentum acceptance $p = 11.0$ GeV/c $\pm\,24\%$ is divided to get finally twelve beam with momentum acceptance of $\pm2\%$.}
	\label{fig:Fig6}
\end{figure*}

More antiprotons are needed, if luminosity is to be increased. To do this a larger momentum spread beam is accepted and then quickly diverted into a dozen beams with smaller momentum spreads. In detail, the antiproton beam coming from the Li lens is dispersed using a 1.5 m long dipole with a magnetic field of 1.8 T.  Particles with high momentum experience less deflection. Once the beam is dispersed, it can be divided by placing an electrostatic septum, and after that two magnetic septum to increase the separation. The electrostatic septum has a thin 0.2\,mm wire plane  minimizing beam loss. The dipoles septa are either 4 or 20\,mm. The beam exiting from the electrostatic septum should have greater separation than the magnetic septum thickness to minimize any interaction with the material. Fig.\,\ref{fig:Fig5} 
presents the basic cell to divide the initial dispersed beam into two:
\begin{itemize}
	\item An initial beam with Landau function momentum distribution (11 GeV/c, 4.2 GeV/c) enters the Li lens and then is spread by a -1.8 T dipole, 1.5 m in length.
	\item An electrostatic septum (ES) divides the beam, which is placed in such a way that half is deflected. The parameters used for the septum are: length = 6 m, Electric Field = $\pm$1.0 MV/m, gap = 0.30 m and septum thickness = 0.2 mm. 
	\item Next to the electrostatic septum a 0.1 T magnetic septum (MS) with septum thickness of 4.0 mm is placed to increase the beam separation. This magnet provides a deflection range of 27-44 mm into the momentum acceptance required.  
	\item A 3.0 m long magnetic septum, 1.0 T, is placed next to the 0.1 T dipole to allow a greater separation between the divided beam. 
	\item To transport the beam a FODO cell (12.5 m) is used. This consists of a focusing quadrupole FQ, a drift space and a defocusing quadrupole DQ. The quadrupoles are 0.66 m in length, aperture radius of 1.0 m and a 2 T/m field gradient.  
	\item The process is repeated to separate the beam into two again, obtaining two beams, and finally each of these beams is separated into three to get the first six beams (Fig.\,\ref{fig:Fig6}).
	\item To obtain the next six beams, the initial half beam, which is was not deflected is transported to be dispersed using a second -1.8 T dipole. Then, the same configuration is used to obtain the other six beams. At the end 12 beams are obtained as is shown in Fig.\,\ref{fig:Fig6}. 
Table \ref{Table2} presents the parameters of the cell which divides the initial beam.
\end{itemize}

\begin{table*}[t]
	\centering
	\caption{Parameters of the basic cell to divide the initial beam.}
		\begin{ruledtabular}
		\begin{tabular}{lccccccc}
			& \textbf{Dipole} & \textbf{FQ} & \textbf{ES} & \textbf{MS1} & \textbf{MS2} & \textbf{DQ} & \textbf{Unit}  \\
			\hline
			Magnetic Field $B$ & -1.8 & - & 0.017 & 0.1  &  1.0 & - & T \\ 
			Field Gradient $G$ & - & 2.0 & - & - & - & -2.0 & T/m \\ 
			Length $L$ & 1.5 & 0.66 & 6.0 & 3.5 & 3.0 & 0.66 & m \\ 
			Radius $R$ & - & 1.0 & - & - & - & 1.0 & m \\  
			Width $w$ & 0.40 & - & 0.35 & 1.0 & 1.0 & - & m \\   
			Septum thickness & - & - & 0.2 & 4 & 20 & - & mm \\
				\end{tabular}
			\label{Table2}
	\end{ruledtabular}
\end{table*}

\begin{figure}[!htb]
	\centering
	\includegraphics*[width=95mm]{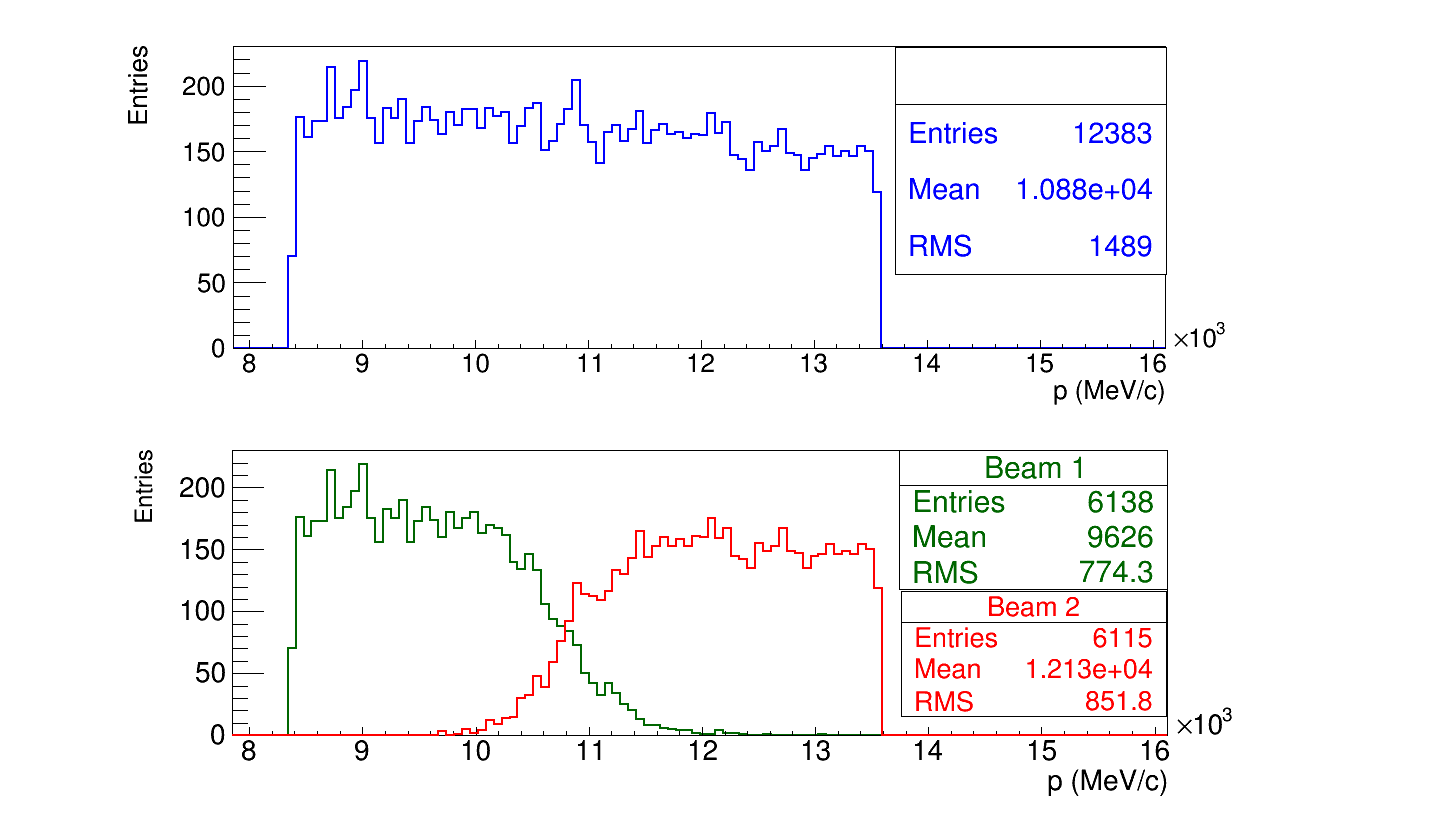}
	\caption{Momentum distribution of the initial beam (upper plot) and after it is divided into two (bottom plot).}
	\label{fig:Fig7}
\end{figure}

\begin{figure}[!htb]
	\centering
	\includegraphics*[width=80mm]{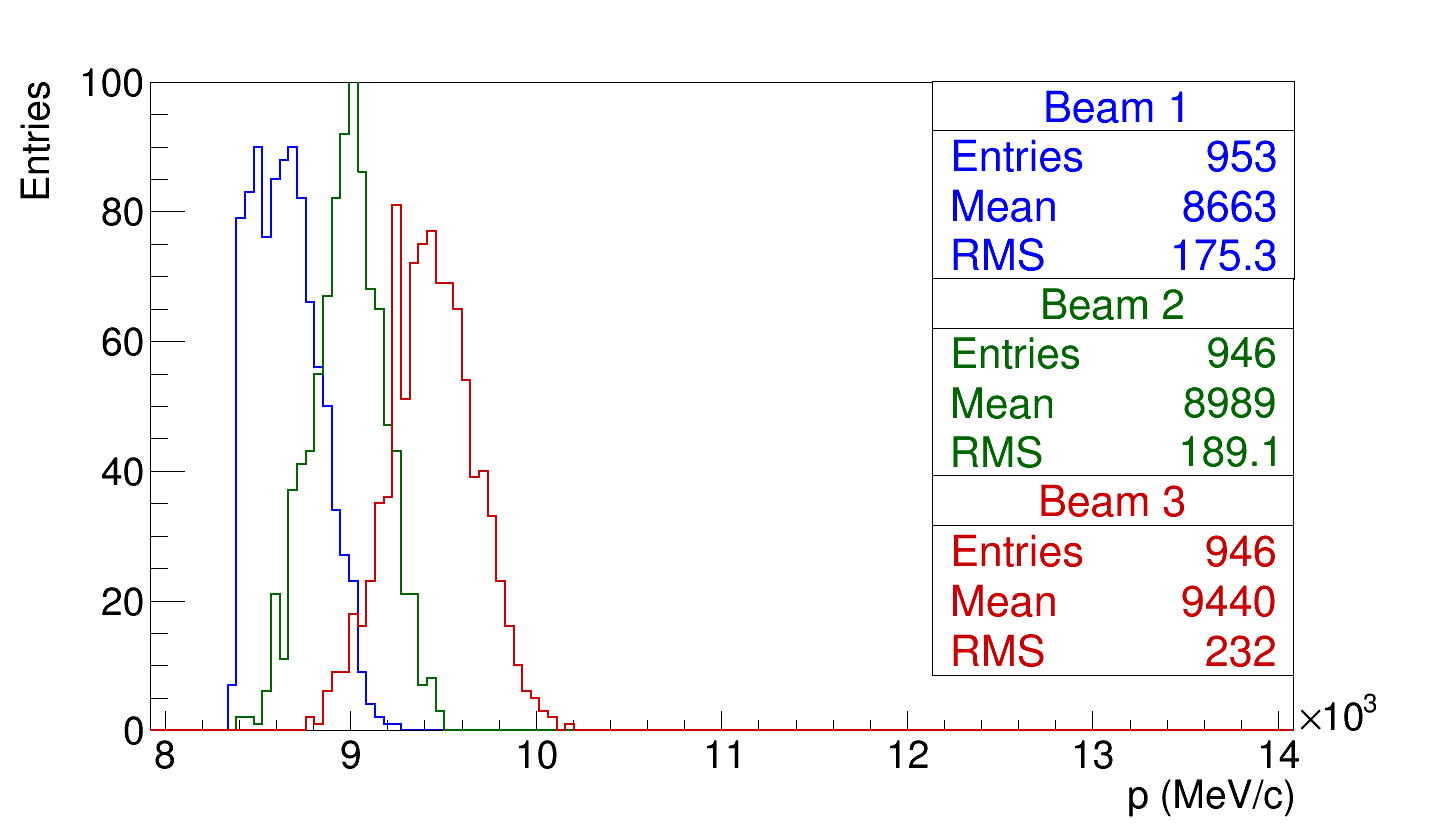}
	\caption{Momentum distribution of the beams 1, 2 and 3.}
	\label{fig:Fig8}
\end{figure}

\begin{figure}[!htb]
	\centering
	\includegraphics*[width=80mm]{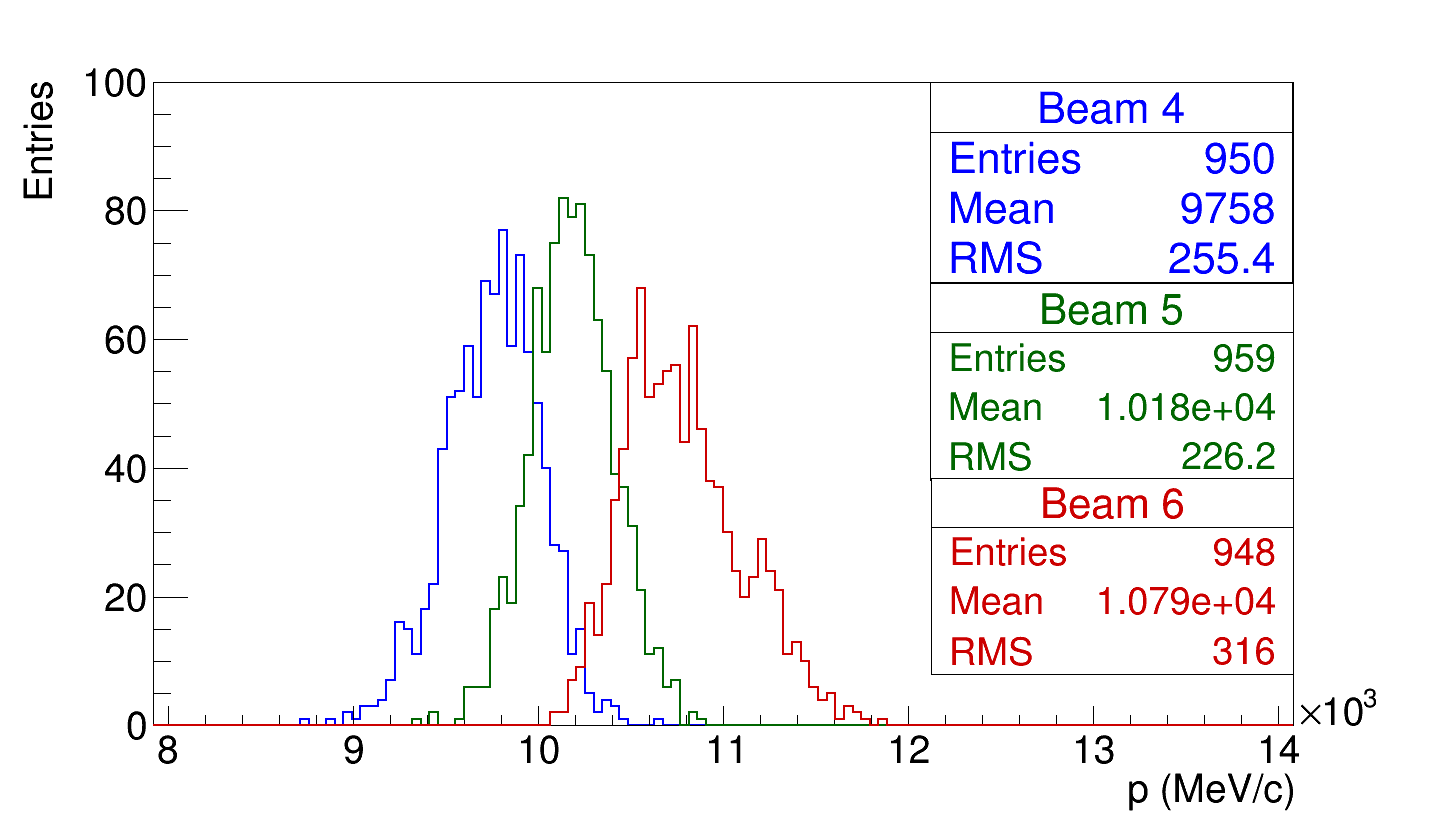}
	\caption{Momentum distribution of the beams 4, 5 and 6.}
	\label{fig:Fig9}
\end{figure}

\begin{figure}[!htb]
	\centering
	\includegraphics*[width=80mm]{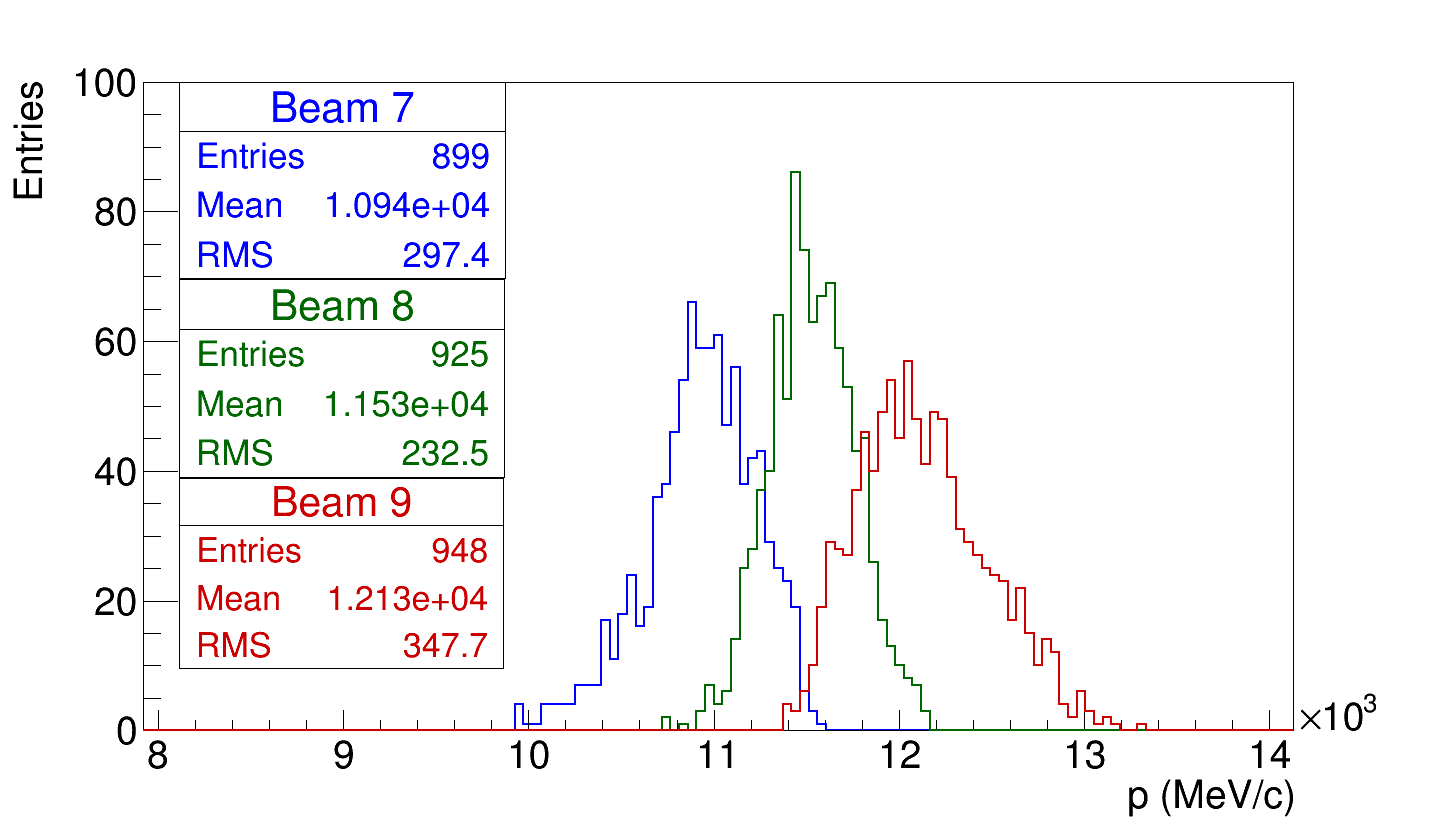}
	\caption{Momentum distribution of the beams 7, 8 and 9.}
	\label{fig:Fig10}
\end{figure}

\begin{figure}[!htb]
	\centering
	\includegraphics*[width=80mm]{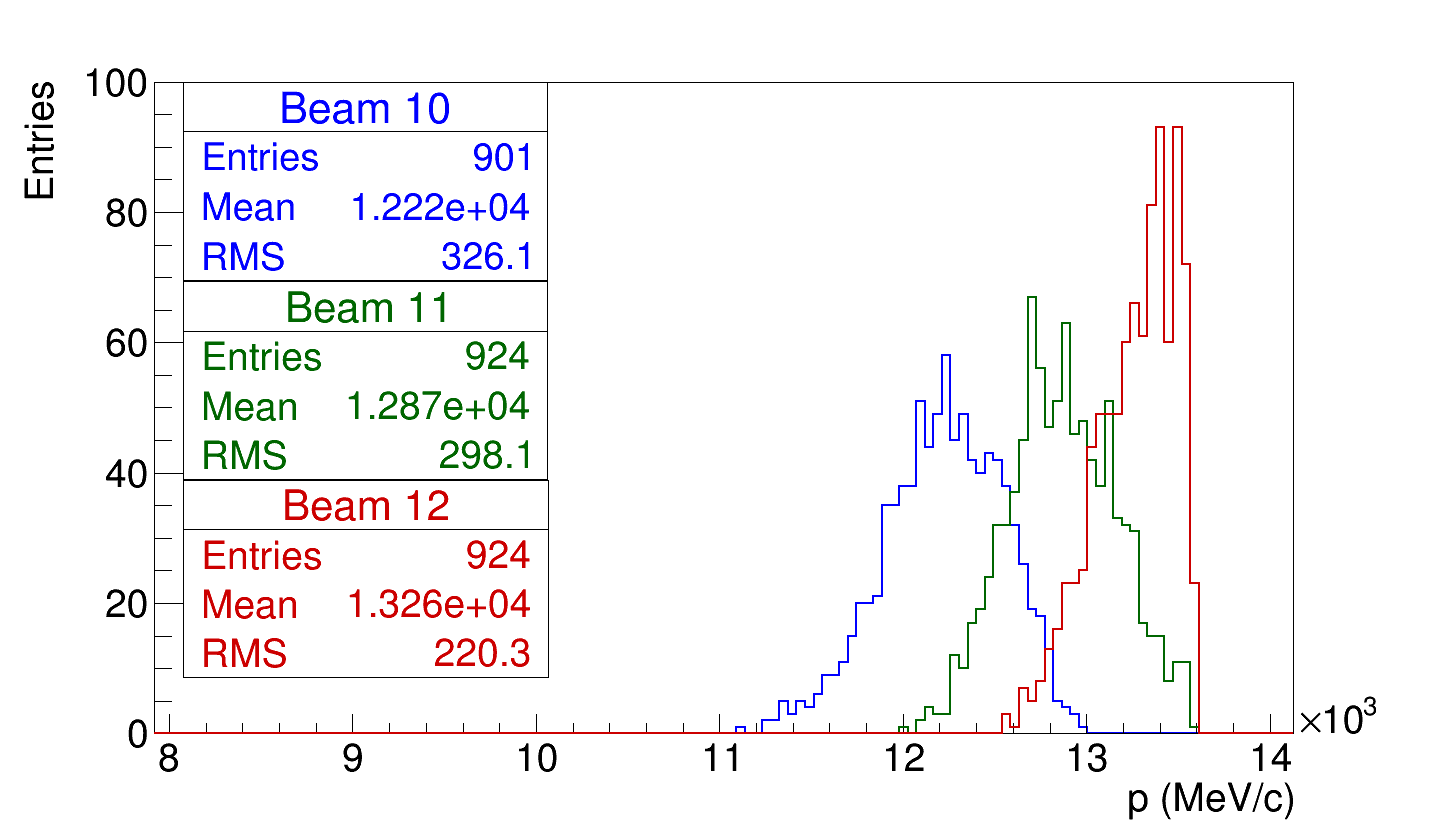}
	\caption{Momentum distribution of the beams 10, 11 and 12.}
	\label{fig:Fig11}
\end{figure}

The momentum distribution of each beam is calculated  using G4beamline as soon as a beam is divided. For example, Fig.\,\ref{fig:Fig7} shows the distribution in momentum of the initial beam (upper plot) and after it is divided into two (bottom plot). 
The momentum distributions of the final 12 beams are shown in Fig.\,\ref{fig:Fig8}, Fig.\,\ref{fig:Fig9}, Fig.\,\ref{fig:Fig10}, and Fig.\,\ref{fig:Fig11}. Table \ref{Table3} presents the mean values of the momentum of each beam together with the width of  each distribution. 

\begin{table}
	\caption{Mean momentum value of the 12 beams obtained.}	
	%	\resizebox{\textwidth}{!}{
	\begin{ruledtabular}
	\centering	
	\begin{tabular}{cccccc}
		& & \textbf{Beam} & \textbf{p (GeV/c)} & &\\ 
		\hline 
		& & 1 & 8.6 $\pm\,2.0\,\%$ & &    \rule{0pt}{9pt} \\		
		& & 2 & 9.0 $\pm\,2.1\,\%$ & &\\
		& & 3 & 9.4 $\pm\,2.4\,\%$ & &\\
		& & 4 & 9.7 $\pm\,2.6\,\%$ & & \\
		& & 5 & 10.2 $\pm\,2.2\,\%$ & &\\
		& & 6 & 10.8 $\pm\,2.9\,\%$ & &\\
		& & 7 & 10.9 $\pm\,2.7\,\%$ & &\\
		& & 8 & 11.5 $\pm\,2.0\,\%$ & &\\
		& & 9 & 12.1 $\pm\,2.8\,\%$ & &\\
		& & 10 & 12.3 $\pm\,2.6\,\%$ & &\\
	    & &	11 & 12.8 $\pm\,2.3\,\%$ & &\\
		& & 12 & 13.2 $\pm\,1.7\,\%$ & &\\		
	
	\end{tabular}
	%	}
	\label{Table3}
	\end{ruledtabular}
\end{table}

\section{Antiproton Cooling System}
In the previous section, it was shown how an initial antiproton beam was dispersed and divided into 12 different momentum channels with a transmission of 89$\%$. Now, because the momentum into the Accumulator rings was 8.9 GeV/c, as well as into the Recycler ring, it is necessary to equalize the central momentum of all the 12 beams to that value. The Debuncher rings ramp the beams up or down to 8.9 GeV/c. At Fermilab, antiprotons were stochastically precooled in the Debuncher ring during 2.2 s, with a normalized transverse rms emittance reduction from 330 to 30 $\mu$m\,\cite{Nagaitsev:2014bha}, then sent to the Accumulator ring to be stochastically cooled and stored. There, the transverse emittance was reduced from 30 to 15 $\mu$m. The stochastic cooling time scales as the number of particles\,\cite{Meer2},
\begin{equation}\label{eq:tao}
\tau\approx N  \times 10^{-8}{\rm{s.}}
\end{equation}
To cool 12$\times$ more antiprotons, 12 independent cooling systems \,\cite{McIntyre} would be implemented as is shown in Fig.~\ref{fig:Fig12}. Each Debuncher ring phase rotates the beam to lower the momentum spread and also ramps the beam central momenta up or down to 8.9 GeV/c. Each Debuncher alternately outputs antiprotons to one of two associated accumulator rings. Each system would have a debuncher/momentum equalizer, which would use RF cavities to reduce the 2\% momentum spread by decelerating fast antiprotons and accelerating slow ones. In addition the central momenta of all 12 channels would be equalized. The Debuncher would alternately feed two accumulator rings. At Fermilab the single Accumulator ring could only cool $25 \times 10^{10} \, \bar{p}$/hr. A second accumulator ring  doubles the time in the deposition orbit and reduces required stack sizes. 
Two accumulator rings can keep up with one 40 x 10$^{10}$ $\bar{p}$/hr Debuncher output rate. In addition, a single electron cooling ring could follow the stochastic cooling. Electrons can cool large numbers of low emittance antiprotons in one ring\,\cite{electron}.

\renewcommand{\arraystretch}{1.1}
\begin{table}[h]
	\caption{Fermilab antiproton cooling stages\,\cite{Gollwitzer}. The normalized rms emittance shown
	comes from muliplying geometric rms emittance by $\beta \, \gamma = p/m = 8.9/0.938 = 9.49.$}
	\smallskip
        \tabcolsep=4mm
        	\begin{ruledtabular}        
	\centering	
	\begin{tabular}{lcc}
	            & Transverse  & Momentum\\
	Stage   &  Emittance   & Spread\\ 
	            & $\mu$m        & MeV/c \\
	            \hline
	 Debucher Entrance  & 330 & $\pm$\,200 \\           
	  After Phase Rotation           &      & $\pm$\,9 \\
	  Debuncher Exit         & 30 & $\pm$\,4.5 \\
	   Accumulator Exit       & 15 & $\pm$\,9\\
	   Recycler Exit             & 2 & $\pm$\,1.8 \\      
		\end{tabular}
	         \end{ruledtabular}
		\label{Cooling}
\end{table}

\subsection{Electron cooling option for $\bar{p}$ stacking}
Twenty four stochastic accumulator rings may be difficult to build and operate. Might all antiproton bunches be stacked in a single electron cooling ring? The 8.9 GeV/c antiproton beam started out with a momentum spread of 2\% or about 200 MeV/c.  
Table~\ref{Cooling} shows the progression of antiproton cooling at Fermilab. Only a small amount of additional stochastic cooling had to occur in the Recycler ring before electron cooling could commence. The Debuncher ring had two longitudinal and four transverse stochastic cooling systems. Antiprotons at Fermilab were cooled from 330 $\mu$m to 30 $\mu$m in the Debuncher ring.
Phase rotation alone in the Debuncher was almost enough to lower the momentum spread to the level needed for electron cooling.

The Fermilab electron cooling ring reduced longitudinal emittance by  a factor of two in 30 minutes. In thirty minutes the system in this paper would produce 9000 bunches, which is roughly the number of antiproton bunches in the machine. They would just need to be coalesced with bunches produced in previous half hour intervals.  The current system might be enough. However, electron cooling is proportional to $1/\gamma^2$.  Reducing the total antiproton energy by a factor of three from 8.94 GeV to 2.98~GeV decreases $\gamma$ by a factor of three and increases the cooling rate by a margin of nine. 

The electric field arising from space charge is given by\,\cite{Lee}
\begin{equation}
E_{\mathrm{s}} = E_{\mathrm{w}} - \frac{e \, g_{0}}{4 \pi \, \epsilon_0 \, \gamma^{\,2}} \frac{\partial{\lambda}}{\partial{s}}
\end{equation}
\noindent 
where $E_{\mathrm{s}}$ and $E_{\mathrm{w}}$ are the electric fields at the beam pipe center and wall, $e = 1.6 \times 10^{-19}$ coulombs, $g_0 = 1 + 2 \ln (b/a)$ is a geometry factor, $a$ is the beam radius, $b$ is the beam pipe radius, $\epsilon_0 = 8.85 \times 10^{-12}$ farads/meter, and $\lambda$ is the antiproton line density.
The Recycler had four RF cavities and  a combined total of accelerating gap voltage  of 2 kV\,\cite{Dey}. This might have to be increased to 36 cavities and 18 kV to control space charge, if the ring energy were lowered by a factor of three.

\subsection{Recycling antiprotons in the collider ring}

\begin{figure}[b!]
	\centering
		\includegraphics*[width=85mm]{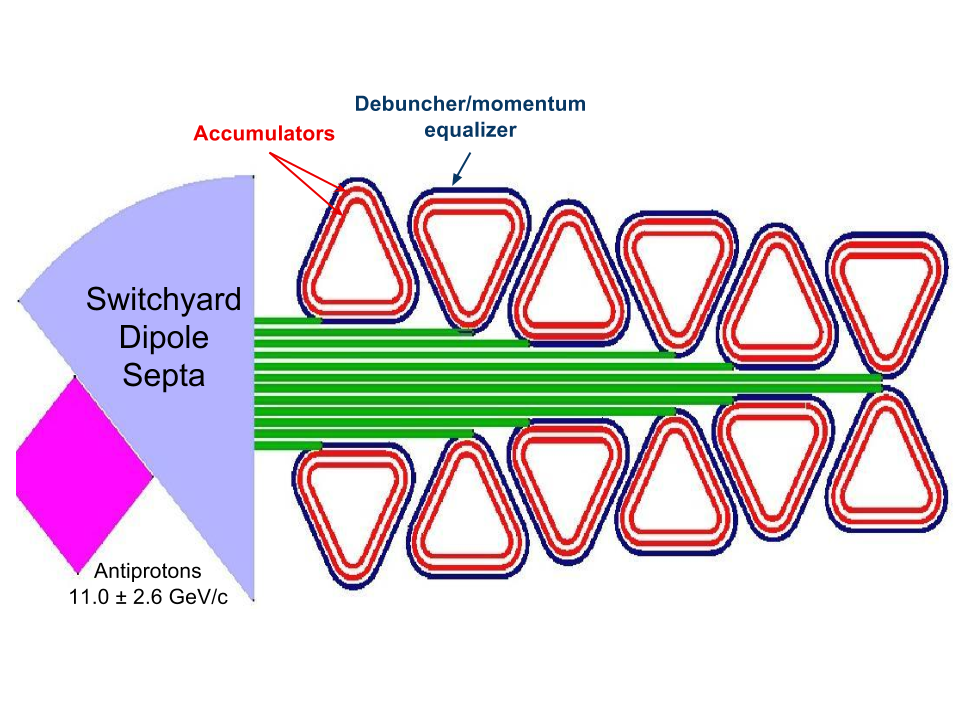}
		\caption{To cool 12x more antiprotons, 12 independent cooling systems would be implemented. 
		Two accumulator rings may be able keep up with one $40 \times 10^{10}\,\bar{p}/hr$ Debuncher output rate.
		Alternatively,  bunch coalescing might be performed with electron rather than stochastic cooling.}
		\label{fig:Fig12}
\end{figure}

Antiprotons are harder to produce than protons. Antiprotons must be cooled. It is appealing to reuse $\bar{p}$\,'s, rather than dumping them at the end of runs. Antiprotons in the collider ring might be recycled without leaving the collider ring. This would increase the availability of antiprotons by about a factor of two. To allow this, the beam energy would have to be occasionally lowered as was done at the CERN S$p\bar{p}$\,S ramping run\,\cite{Albajar} in 1985, where the S$p\bar{p}$\,S collision energy was ramped from 200 to 900 GeV (100 to 450 GeV/beam). Continuous (trickle charge) injection improved integrated luminosity at PEP-II\,\cite{Seeman}.

New antiproton bunches would be placed next to old bunches using snap bunch coalescence. In this process two radio frequency (RF) systems operate with different frequencies. The bunches are rotated for 1/4 of a synchrotron oscillation period at low frequency and then captured at high frequency\,\cite{coalescing}.

Finally, new and old antiproton bunches would be completely coalesced with synchrotron damping. Synchrotron damping decreases transverse emittance and compensates for the loss of antiprotons  from collisions so that luminosity can be maintained 
during a run\,\cite{McIntyre2015}.

\begin{figure*}[t!]
	\begin{center}
		\includegraphics[width=6.2in]{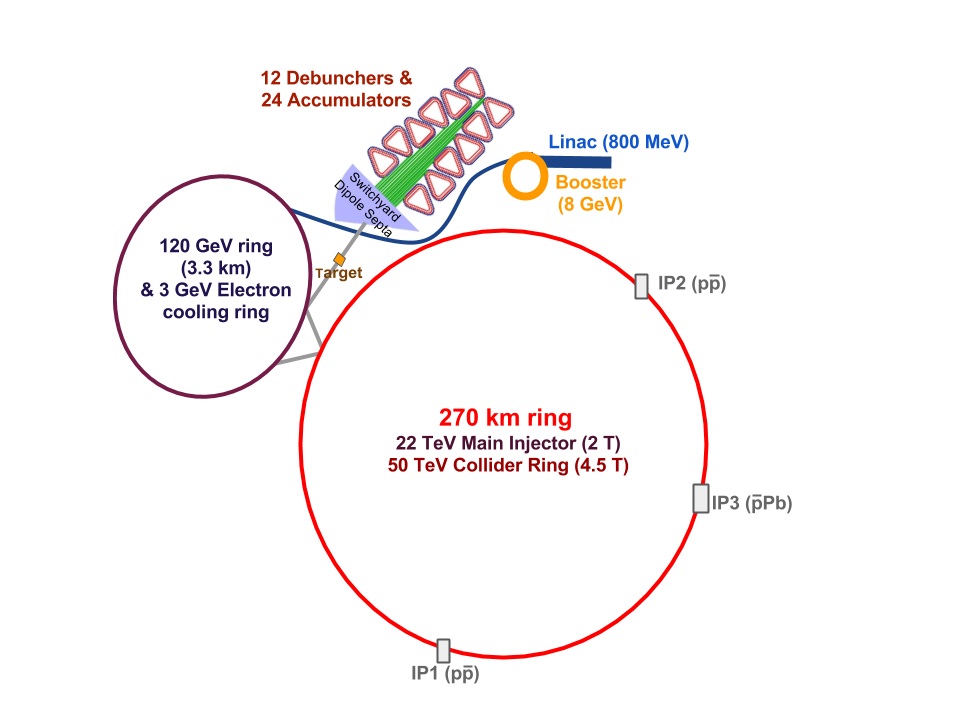}
		\caption{100 TeV proton-antiproton collider (not to scale). An intermediate energy ring between 120 GeV and 22 TeV is not shown.
		Bunch stacking in the single electron cooling ring might replace the 24 Accumulator rings. A bit of additional transverse cooling would have to be coaxed 
		out of the  Debuncher rings to allow electron cooling.}		
		\label{fig:Fig6-1}
	\end{center}
\end{figure*}
	   
\section{Collider Layout}
For the construction of the 100 TeV $p\bar{p}$ collider tunnel, two possible rock strata are considered: Fermilab dolomite\,\cite{VLHC} and Texas chalk\,\cite{Peter}. Fermilab has the advantage of existing infrastructure and Texas has lower tunneling costs. Figs.\,\ref{fig:Fig12} and\,\ref{fig:Fig6-1} show the configuration proposed. The components are described below:
\begin{itemize}
	
	\item An upgraded 800 MeV super-conducting Linac\,\cite{Derwent:2015qfa}, which Fermilab has proposed to provide megawatt proton beams for muon and neutrino experiments. The 190 m long Linac would accelerate  H$^{-}$ ions to an energy of 800 MeV before  passing them through a thin carbon foil to remove the electrons.
	This allows charge exchange injection into the Booster and avoids a kicker magnet, as was the case with 400 MeV protons. More 800 than 400 MeV protons can be injected into the Booster. 
	\item The Booster accelerates the 800 MeV protons to a kinetic energy of 8 GeV. The Booster would run at 15 Hz.
	\item The 120 GeV ring receives the 8 GeV protons to be accelerated to 120 GeV, which are sent to the antiproton source to produce the antiprotons.
	\item For antiproton production, a Fermilab-like antiproton source is adapted to the new collider. This will collect 12$\times$ more antiprotons with the switchyard dipole septa, where the antiprotons are dispersed and separated into 12 different momentum channels. For antiproton stochastic cooling 12 sets of rings are implemented.
	\item The electron cooling ring provides additional cooling to the antiprotons coming from the antiproton source. The 20 m long electron cooling system is inserted inside the 3 GeV ring, which would be 3.3 km in circumference.  Ideally, the electron cooling ring will do the bunch stacking and stochastic stacking rings will not be needed.
	\item In the 120 GeV ring, both protons and antiprotons are accelerated to 120 GeV before transfer to the 22 TeV Injector.  An intermediate energy ring may also be useful.
	\item The 22 TeV Injector accelerates the protons and antiprotons to 22 TeV. This energy would be reached using 2 T magnets in the 270 km ring tunnel.
	\item Finally, the 50 TeV collider ring will accelerate the protons and antiprotons from 22 TeV to 50 TeV to collide them with a 100 TeV energy center of mass. The 50 TeV energy would be reached using 4.5\,T magnets in a 270 km circumference ring. Both 22 and 50 TeV rings share the same tunnel.
\end{itemize}

The relatively  inexpensive 2\,T superferric magnets ring would be built first and used as a collider. Ultra low carbon steel would be employed\,\cite{Laeger}. These would be H frame, iron dominated magnets with superconducting  MgB$_2$ cable in conduit at a temperature of 25\,K. At today's \$5 per kiloamp-meter cost, the total conductor price would be  approximately \$150M.
Cooling would be with neon, which costs \$100 per liquid liter. Liquid neon has a heat of vaporization of 1.71 kilo\,-Joule/mole, much larger than 0.083 kilo\,-Joule/mole for helium. The MgB$_2$ would be in the magnet fringe field at a fraction of a Tesla.
CERN plans to use MgB$_2$ as magnet leads for superconducting Nb$_3$Sn quadrupoles\,\cite{CERN-MgB2}.  

The 4.5 T NbTi magnets would be an upgrade. Collisions would include $p \bar{p}$, $\bar{p}\,Pb$, and asymmetric (2\,T and 4.5\,T rings) $Pb\,Pb$. Lepton colliders could also share the tunnel\,\cite{Summers, leptons}. The 2\,T injector ring would horizontally bypass $p \bar{p}$ detectors.

\begin{figure}[b!]
	\begin{center}
	\hspace*{-15mm}
		\includegraphics[width=5.0in]{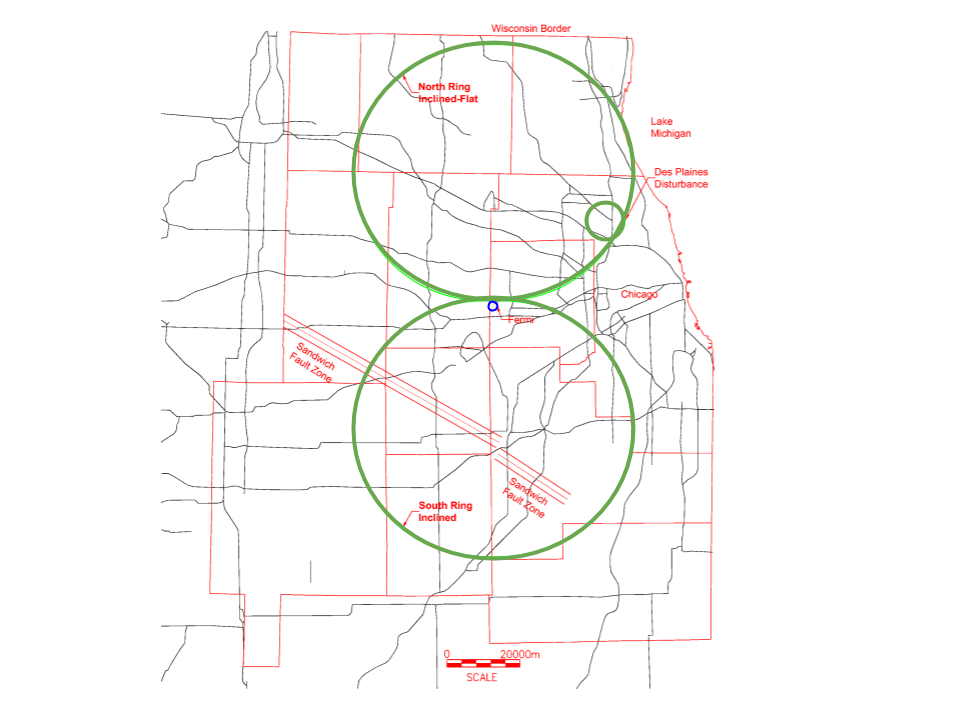}
		\caption{Layout of 233 km ring options, one north and one south of Fermilab\,\cite{VLHC} in Illinois.}
		\label{Fermilab-rings}
	\end{center}
\end{figure}

In Illinois, a large ring\,\cite{VLHC} might be connected to Fermilab as shown in Fig.\,\ref{Fermilab-rings}. An engineering study of the Illinois tunnel has been performed.

\begin{figure}[h!]
	\begin{center}
	\hspace{-8mm}
		\includegraphics[width=3.6in]{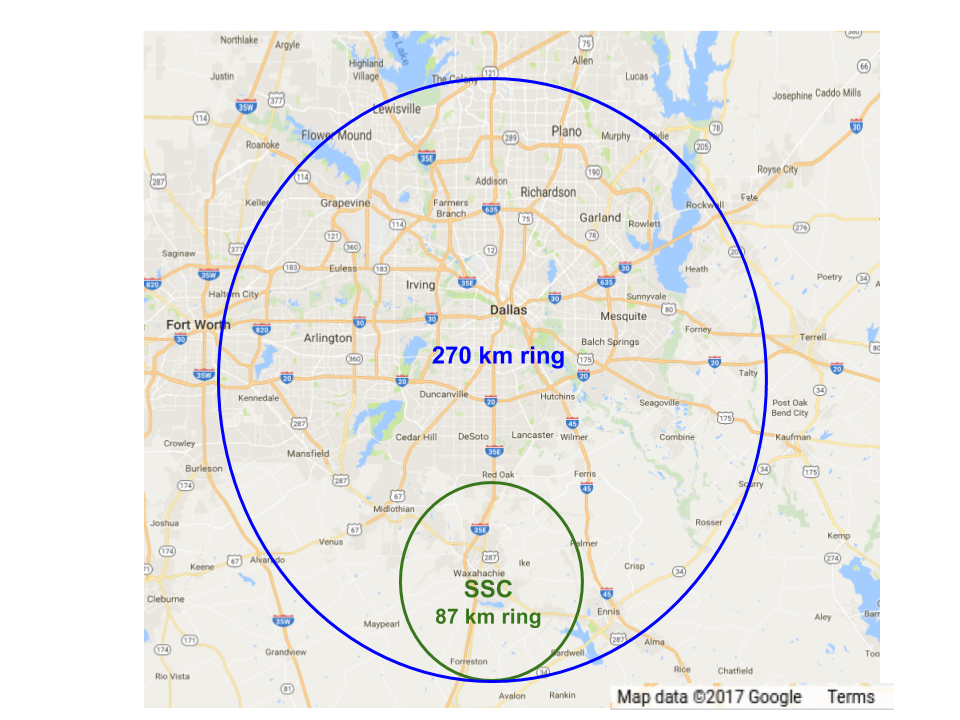}
		\caption{Layout of the 270 km ring around Dallas, Texas\,\cite{Peter}.}
		\label{fig:Fig6-2}
	\end{center}
\end{figure}

In Texas, the 270 km ring might be connected to the partially completed SSC (Superconducting Super Collider)\,\cite{SSC} ring tunnel. Of its initial 87 km design, 45\% was bored. This geographical zone has the advantage of a homogeneous soft rock composition allowing rapid and cheaper tunnel boring, which is a way to  to reduce costs. Fig.\,\ref{fig:Fig6-2} shows the 270 km ring on a map. 

\begin{figure}[h!]
	\begin{center}
		\includegraphics[width=3.2in]{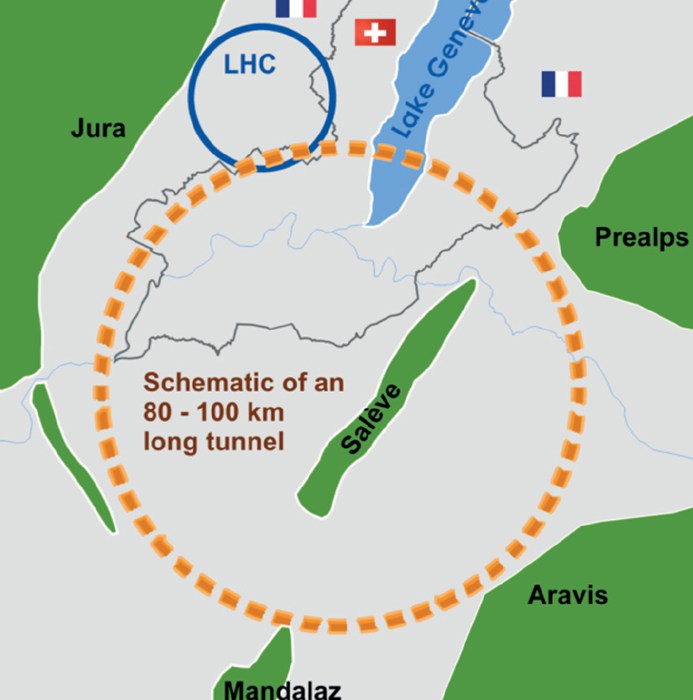}
		\caption{80-100 km tunnel to host a 100 TeV pp collider at CERN in the Geneva valley. Image credit: CERN.}
		\label{fig:Fig6-3}
	\end{center}
\end{figure}

A 270 km ring would be very difficult  to bore at CERN due to the costs of tunneling under the French Jura mountains. The rock under the Jura is very hard. The construction of a 100 TeV collider FCC (Future Circular Collider) of 80\,-100 km ring in the Geneva area as shown in Fig.\,\ref{fig:Fig6-3} may be possible. This would be for both  positron-electron ($e^{+}e^{-}$) and proton-proton colliders. For this 80\,-100 km ring, geologic conditions are being evaluated\,\cite{Kenyon}. 

\section{Tunneling}
Tunneling cost depends on  geology. Table\,\ref{tunneling} shows the cost per meter for boring 4 m diameter tunnels in three different locations. These values are taken from M. Breidenbach and W. Barletta, ESS-DOC-371\,\cite{Breidenbach} to estimate the total cost of a 270 km tunnel. At CERN the tunnel is limited to 80\,-100 km due to the French Jura mountains. A tunnel larger than 100 km would cost much more than \$39,000/m. For this reason the total cost for a 270 km tunnel at CERN is not presented in the table. Now, comparing the total cost for tunneling at Fermilab and Texas, the cost difference is a factor of 2.5. Texas chalk is easier to  bore than Illinois dolomite. 

\begin{table}[h]
	\centering
	\caption{Comparison between tunneling cost for three different locations considered for a 270 km collider ring\,\cite{Breidenbach}.} 
	\vskip .1 in
	%	\resizebox{\textwidth}{!}{
	\begin{ruledtabular}	
	\begin{tabular}{lcc}
		 & Cost/m & 270 km tunnel \\  
		 \hline
		CERN (Molasse/limestone) & \$39,000 & 100 km limit\\
		FERMILAB (Dolomite) & \$15,000 & \$4 billion \\
		Texas(Chalk/marl) & \$6,000 & \$1.6 billion \\
	\end{tabular}
	\end{ruledtabular}	
	%	}
	\label{tunneling}
\end{table}

The LEP tunnel construction project took 4 years, beginning in 1985 and finishing in 1989. Three tunneling machines were used to bore the 4 m diameter tunnel. The average bore rate per tunneling machine was approximately 5 meters/day. The three tunnels that compose the channel tunnel between France and the UK are each 50 km long. The project  took 6 years using eleven tunneling machines. All this information is shown in Table\,\ref{tunneling-time}, along with the calculation of the volume of rock removed. The channel project consisted of two 8 m diameter tunnels plus one 4 m diameter tunnel. From this, we can estimate that the rock removed was 0.025 million m$^{3}$/year per machine  for LEP and 0.085 million m$^{3}$/year per machine for the Channel Tunnel. For a 270 km tunnel, 3.4 million m$^{3}$ of rock would have to be removed. 

However, rock in Texas is faster to bore; 40 m/day based on SSC tunneling rates\,\cite{McIntyre2015}. Thus, it would take about 4.6 years using 4 tunneling machines, even better, 3.7 years using 5 tunneling machines. Fermilab has  relatively constant dolomite layers. For a 4 m diameter tunnel the average advance rate for a tunnel boring machine is about 20 m/day\,\cite{Foster1996}. Thus, it can be estimated that using 8 boring machines the tunneling time is roughly 4.6 years.
\begin{table}[h]
	\centering
	\caption{Tunneling-time estimate.} 
	\smallskip
	%	\resizebox{\textwidth}{!}{	
	\begin{ruledtabular}	
	\begin{tabular}{lcccc}
		& Length & Volume of rock & Time & Tunneling \\ 
		& (km) & (million m$^{3}$) & (years) & Machines \\ 
		\hline
		LEP & 27 & 0.3 & 4 & 3\\
		Channel Tunnel & 3$\times$50 & 5.6 & 6 & 11\\
		Tunnel (Illinois) & 270 & 3.4 & 4.6 & 8\\
		Tunnel (Texas) & 270 & 3.4 & 3.7 & 5\\ 
	\end{tabular}	
	\end{ruledtabular}	
	%	}
	\label{tunneling-time}
\end{table}

\section{Main Dipole Magnets}

\begin{figure}[h!]
	\begin{center}
		\includegraphics[width=3in]{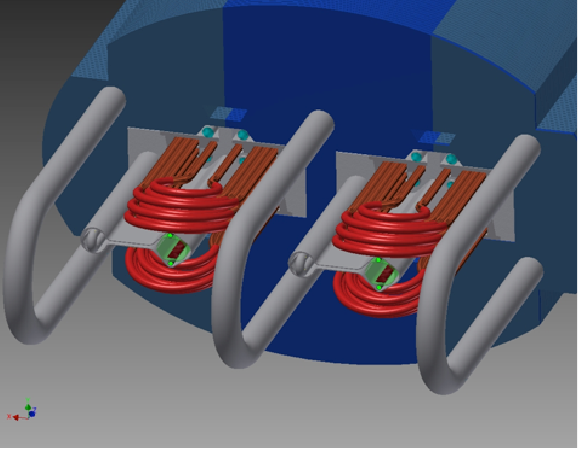}
		\caption{4.5 T dual dipole magnet \cite{McIntyre2015}. Only one bore would be needed for a $p \bar{p}$ collider.}
		\label{fig:Fig6-5}
	\end{center}
\end{figure}

\begin{figure}[h!]
	\begin{center}
		\includegraphics[width=3in]{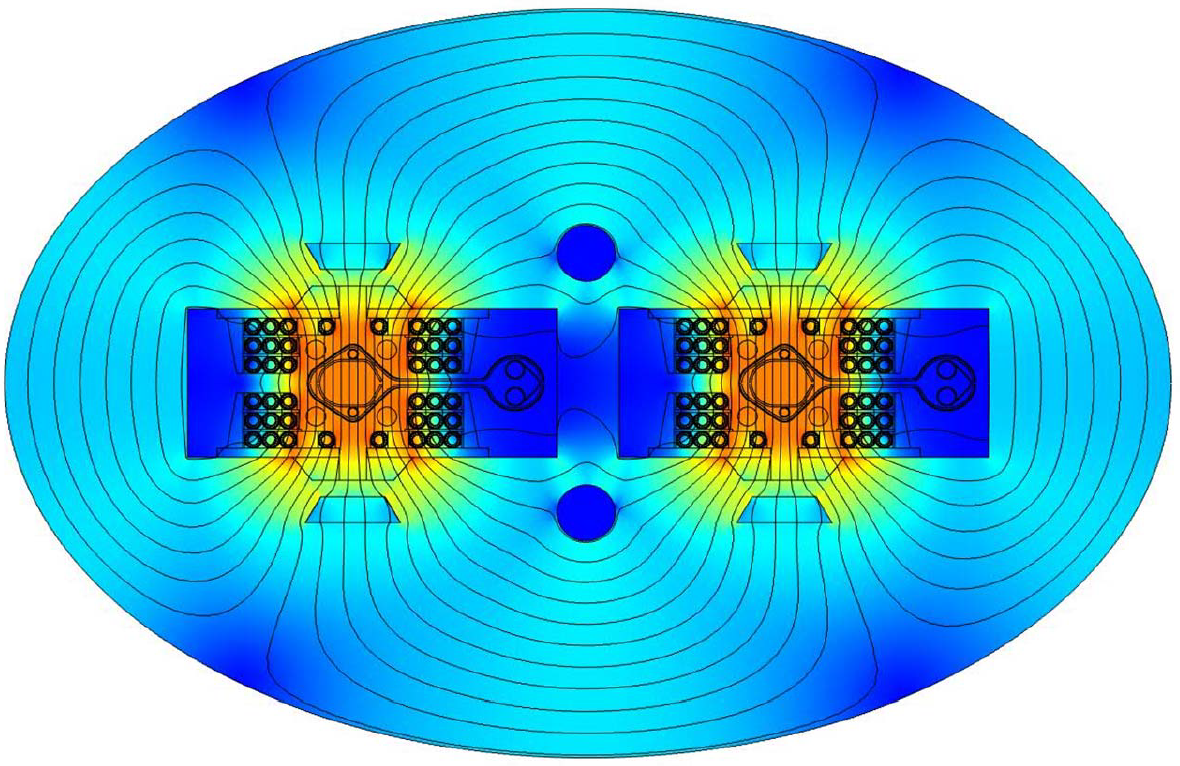}
		\caption{4.5 T dual dipole magnetic design \cite{McIntyre2015}. Only one bore would be needed for a $p \bar{p}$ collider.
		The bore size is approximately 25 by 30 mm.
		}
		\label{fig:Fig6-6}
	\end{center}
\end{figure}

In a collider the dipole magnets represent a large budget item. The main LHC magnets are located in the 27 km main ring with a packing fraction of 66\%, which gives a bending radius is 2.8 km. According to the equation, $\mathrm{\rho [m]=p[GeV/c]/0.3B[T]}$, the magnetic field is $\sim$8.3 T for an energy of 7 TeV. These twin bore magnets are made of Niobium-Titanium (NbTi) superconducting material. Two layers of NbTi cable are distributed to form a $\cos\,\theta$ structure. The dipole is 14.3 m long, it operates at 1.9 K and the total current for a 8.3 T magnetic field is 11.8 kA. The LHC required 1232 main dipoles, which represented a total cost of \$660 million. The cost of each dipole magnet was \$0.5 million. Futhermore, the LHC dipole magnet cost around 3 times more than its superconductor.
A 100 km $pp$ collider with $E_{CM}=100$ TeV requires 16 T magnets, which would need Nb$_{3}$Sn material and its production is still under study\,\cite{Bottura}. A better balance between magnet and tunneling costs may be found at lower magnetic fields.
Stored magnetic field energy decreases linearly with ring circumference. A 100 TeV proton-proton collider in a 270 km tunnel around Dallas, Texas using 4.5 T dipole magnets has been proposed\,\cite{Peter}. These magnets use superconducting NbTi cable in conduits as is shown in Fig.\,\ref{fig:Fig6-5}. Twenty turns of cable make up each dipole winding. The magnet  operates at 4.5 K. These 4.5 T magnets use about half as much NbTi conductor per Tesla/meter as 8 T $\cos \theta$ LHC magnets. Fig.\,\ref{fig:Fig6-6} shows the magnetic configuration of the dipole.

For the 100 TeV $p \bar{p}$ collider, the energy of the Injector would be 22 TeV. Using 2 T superferric, iron dominated magnets, with a packing fraction of 86\%. The Injector Ring would share the same 270 km tunnel as the collider ring. The injector could offer 44 TeV collisions in a first stage. These superferric magnets need  minimal current because the magnetic field just has to be generated inside the gap. The iron does limit the maximum magnetic field generated 
to 2\,T. 

The high luminosity, 100 TeV $p\bar{p}$ collider resides in the  270 km tunnel. With a packing fraction of (86\%), the bending radius would be 37 km, which would requires $\sim$4.5 T magnets for a 50 TeV energy beam. Because of the advantages of the 4.5 T dipole magnets as described above, these are proposed for the 270 collider ring with the difference that only a single bore is needed, instead of a dual bore as required for a proton-proton collider. It is important to note that with respect to the LHC collider, the energy is increased 7 times using only a factor of two  more NbTi superconductor. 

\section{Parameter Calculation}
Table\,\ref{Colliders-parameter} lists the main parameters for the Tevatron, the LHC, a 100 TeV ($pp$) Future Circular Collider FCC-hh \cite{Barletta}, and this 100 TeV $p\bar{p}$ collider. The calculations of the parameters for the 100 TeV $p\bar{p}$ collider are explained below: 

\renewcommand{\arraystretch}{1.15}
\begin{table*}[t]
%\begin{sidewaystable}
	\centering
	\caption{Parameters for the Tevatron, the LHC, the Future Circular Collider FCC-hh, and the 100 TeV $p\bar{p}$ proposed here.}
	%	\resizebox{\textwidth}{!}{	
	\begin{ruledtabular}	
	\begin{tabular}{lccccc}
			\textbf{Collider Parameters} & \textbf{Tevatron} & \textbf{LHC} & \textbf{FCC-hh} & \textbf{100 TeV $\boldsymbol{\mathrm{p\bar{p}}}$} & \textbf{Unit}  \\
			\hline
			Luminosity $(L)$  \rule{0pt}{11pt} & 3.4 x 10$^{32}$ & 1.0 x 10$^{34}$ & 5.0 x 10$^{34}$ &  1.0 x 10$^{34}$ & $\mathrm{cm^{-2}s^{-1}}$  \\  
			Energy Center of Mass $(E_{cm})$ & 1.96  & 14  & 100  & 100 & $\mathrm{TeV}$   \\  
			Magnetic Field $(B)$ & 4.3  & 8.3  & 16  &  4.5 & $\mathrm{T}$   \\    
			Packing fraction & 0.77 & 0.66 & 0.66 & 0.86 & \\
			Circumference $(C)$ & 6.28  & 27   & 100 & 270  & $\mathrm{km}$   \\ 
			Bending Radius ($\rho$) & 760 & 2801 & 10416 & 37040 & m \\
			Revolution Frequency $(f_{0})$ & 0.048 & 0.01  & 0.003  & 0.0011  & $\mathrm{MHz}$   \\
			Collision Frequency $(f)$ & 1.7   & 40  & 40 & 12  & $\mathrm{MHz}$   \\   
			Lorentz Gamma Factor $(\gamma)$ & 1044   & 7460 & 53304  & 53304   &    \\ 
			Number of Bunches $(N_{B})$ & 36   & 2808    & 10600 & 10800  &   \\ 
			Number of Protons/Bunch $(N_{p})$ & 29 x 10$^{10}$ & 11.5 x 10$^{10}$   & 10 x 10$^{10}$ & 20 x 10$^{10}$ &   \\ 
			Number of Antiprotons/Bunch $(N_{a})$ & 8 x 10$^{10}$ &   &  & 0.32 x 10$^{10}$  &   \\ 
	     	Total/Inelastic Cross Section  & 81.9 / 61.9 & 111 / 85  & 153 / 108 & 153 / 108 & mb \\ 
			Events per Bunch Crossing  & 12 & 27  & 170 & 90 &   \\ 
			Norm. RMS Transverse Emittance $(\varepsilon_{N})$ & 3.0 (protons) & 3.75  & 2.2  & 2.2 (protons) & $\mathrm{\mu m}$  \\ 
			& 1.5 (antiprotons) &  &  & 1.5 (antiprotons) & $\mathrm{\mu m}$   \\
			Betatron Function at IP $(\beta^{*})$ & 0.28  & 0.55   & 1.1 & 0.14 & $\mathrm{m}$   \\ 
			Beam Size at IP $(\sigma)$ & 33 (protons)  & 16.6  & 6.8 & 2.4 (protons)  & $\mathrm{\mu m}$   \\ 
			& 29 (antiprotons)  &  &   &  2.0 (antiprotons)  & $\mathrm{\mu m}$   \\ 
			Beam-Beam Tune Shift per IP $(\xi)$ & 0.006 (protons)  & 0.003   & 0.005 & 0.0003 (protons) &  \\ 
			& 0.012 (antiprotons)  &  &  & 0.011 (antiprotons) &   \\ 
			Number of IPs $(N_{IP})$  & 2 & 4 & 2 & 3 &   \\ 
			SR per meter  & 0.00015  & 0.13 & 29 &  2.2 & W/m \\
			Energy loss per turn $(U_{0})$ & 0.0000095  & 0.0067  & 4.6 & 1.3   &  $\mathrm{MeV}$   \\ 
			Longitudinal  Damping Time $(\tau_{\varepsilon})$ & 305  & 13 & 0.5 & 4.8  & $\mathrm{h}$   \\ 
			Transverse  Damping Time $(\tau_{x})$ & 610  & 26  & 1.0 & 9.7 & $\mathrm{h}$   \\ 
		\end{tabular}		
		\end{ruledtabular}			
		%	}
		\label{Colliders-parameter}
%\end{sidewaystable}
\end{table*}

\begin{itemize}
\item The luminosity is 29 times higher than the Tevatron. 

\item The center of mass energy would be 100 TeV, the same as the FCC-hh collider and about 50 times the Tevatron energy.

\item No new technology is required for the 4.5\,T dipole magnets. They are based on NbTi. 

\item A 270 km ring is needed in order to use 4.5 T magnets. A 100 km ring requires 16 T magnets, which are challenging.

\item The revolution frequency is calculated considering that the particles move almost  at the speed of light. The frequency is inversely proportional to the ring circumference. Luminosity is directly proportional to the revolution frequency. 

\item The collision frequency is determined by the number of bunches inside the ring, where the distance between bunches 
is calculated. The bunches  move almost at the speed of light.

\item The Lorentz gamma factor is calculated dividing the beam energy by the particle mass. The luminosity is proportional to this factor.

\item 
The number of antiprotons and the luminosity is fixed. More bunches mean fewer antiprotons per bunch, which decreases the number of events per beam crossing. But to keep the luminosity constant, the number of protons per bunch must remain fixed even with more bunches. This is increases the amount of synchrotron radiation deposited into magnets. Nonetheless, the events per beam crossing and the synchrotron radiation per meter are both lower in the $p \bar{p}$ machine than in the FCC\,-hh machine. 

\item The number of events per bunch crossing equals the cross section times luminosity divided by the collision frequency.

\item The proton beam emittance in the $p \bar{p}$ is taken to be the same as in the FCC\,-hh machine.

\item The beam radius at the interaction point (IP) is given by $\sqrt{\rule{0pt}{8pt} \epsilon_N \beta^* / (\beta \, \gamma)}$, where   $\beta \, \gamma$ = p/m.
 
\item The beam-beam tune shift is given by $\Delta \nu = \xi = N r_p / 4 \pi \epsilon_N$\,\cite{Wu}, where $r_p$ is the classical proton radius. The proton tune shift is caused by the antiproton bunches and vice versa.

\item Two interaction regions are for $p \bar{p}$ detectors and one interaction region is for a $\bar{p} Pb$ detector.

\item The synchrotron radiated power is  calculated using  equation \ref{power}. 

\item 
Energy loss per turn is:  $U_0 = q^2 \beta^3 \, \gamma^4 / (3 \epsilon_0 \rho)$. The longitudinal damping time equals the orbital period times E/U$_0$. The transverse damping time is twice the longitudinal damping time. 
\end{itemize}

\section{Inner Quadrupole System}

\begin{figure}[h]
	\begin{center}
		\includegraphics[width=3.6in]{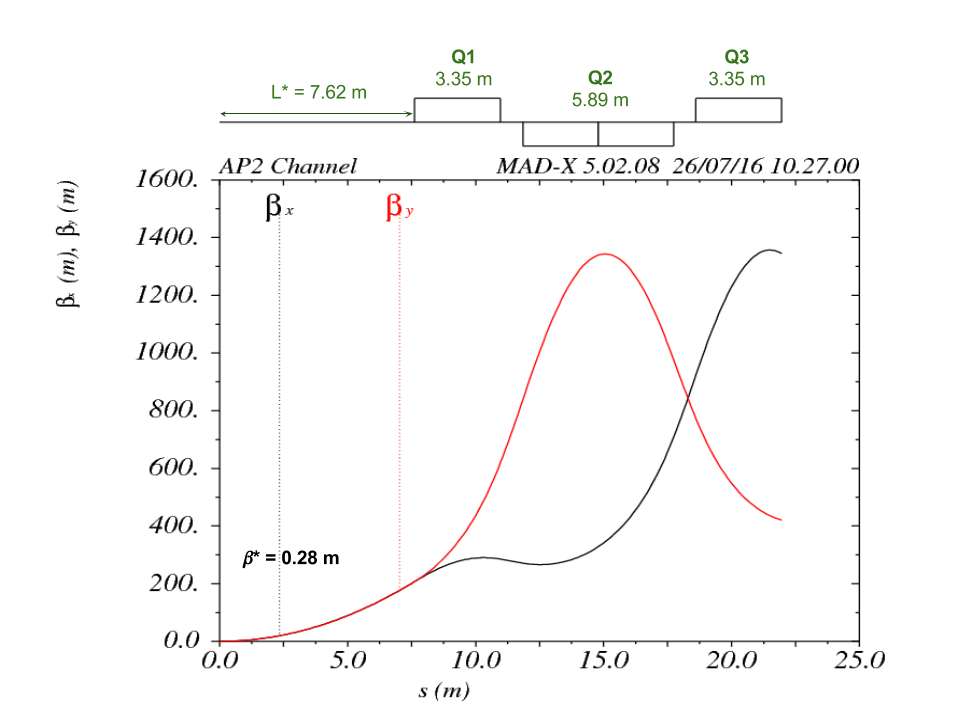}
		\caption{Tevatron D0 interaction region beta functions.} 
		\label{fig:Fig6-8}
	\end{center}
\end{figure}

The inner quadrupole system provides the final focusing of the beams at the collision point. Fig.\,\ref{fig:Fig6-8} shows the Tevatron 
inner triplet quadrupole which focuses the beam at the interaction point (IP). The figure also shows the beta function plot which is related to the beam size at the IP. At the IP, $\beta$* = 28 cm and the quadrupole gradients are 141 T/m for Q1 and Q3, and 138 T/m for Q2.

\begin{figure}[h]
	\begin{center}
		\includegraphics[width=3.4in]{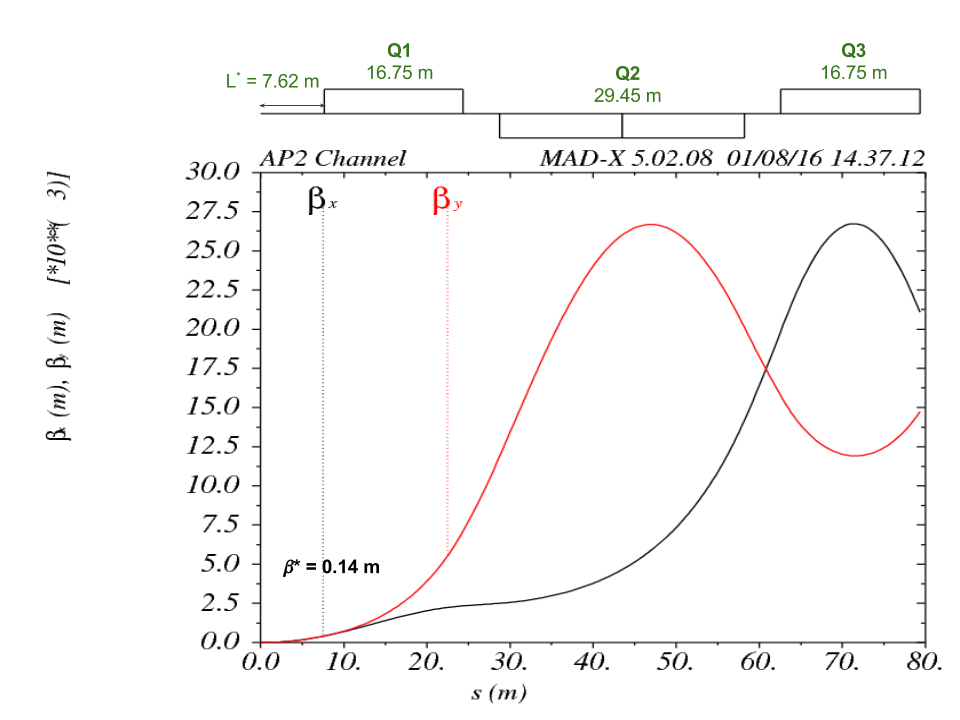}
		\caption{Beta functions plots for the 100 TeV $p\bar{p}$ collider interaction region.} 
		\label{fig:Fig6-9}
	\end{center}
\end{figure}

The Tevatron and LHC have used 8\,T poletip field NbTi quadrupoles, which could be replaced with new technology 13\,T pole tip field Nb$_{3}$Sn quadrupoles\,\cite{IR}. The Tevatron triplet quadrupole system is taking as reference to determine the new parameters for a 100 TeV collision energy. This is simulated using Mad-X\,\cite{Mad-X}, where the $\beta$* value is fixed and the quadrupole lengths are varied in order to get $\beta_{x,max}$=$\beta_{y,max}$ in the beta functions plot. Because a smaller $\beta$* will allow higher luminosity, $\beta^*$ is chosen to be half of the  value at the Tevatron IP.
In order to keep the same distance from the interaction point to the quadrupole Q1 ($L^{*}$) and  to keep $\beta$*= 14 cm, the quadrupole length ($l$) and the separation between the quadrupoles ($a$) are increased by a factor of 5. Fig.\,\ref{fig:Fig6-9} shows the inner triplet quadrupole system scaled, where $\beta_{x,max}$=$\beta_{y,max}$ = 27 km. Also, to get that optimization the field gradients of the quadrupoles are fixed to be 605 T/m for Q1 and Q3, and 354 T/m for Q2. 

The maximum beam size can be calculated using
\begin{equation} \label{eq:size}
	\sigma_{(x,y)max}=\sqrt{\frac{\beta_{(x,y)max}\varepsilon_{N(x,y)}}{\beta_{rel}\gamma}}
\end{equation}
with $\varepsilon_{N(x,y)}$ being the normalized transverse emittance, $\beta_{rel}$=p/E, and $\gamma$=E/m is the Lorentz gamma factor. Using a normalized transverse emittance of $\varepsilon_{N(x,y)}$ = 2.25 $\mu$m, a value of 1.1 mm is obtained for the beam size. The quadrupoles field aperture should be around 10\,$\sigma_{max}$\,\cite{Feher:1996rr} to be large enough for the beam, and a factor of 2 for field quality could be added. This leads to  a 44 mm bore diameter quadrupole. In Table\,\ref{Table5} are listed the inner triplet quadrupoles parameters, corresponding to the Tevatron, the LHC, and the 100 TeV $pp$\,\cite{Martin:2015roa, Luis} and $p\bar{p}$ colliders. There, it can be observed the quadrupole field gradients are higher than those required for the $pp$ collider.  However, the quadrupole bore sizes are smaller and the 13\,T pole tip field limit is not exceeded. 

\tabcolsep=0.5mm		
\begin{table}[t!]
\renewcommand{\arraystretch}{1.15}
	\centering
	\caption{Inner Triplet Quadrupole parameters}
	\label{Table5}
	\begin{ruledtabular}
%	\resizebox{\textwidth}{!}{
		\begin{tabular}{cccccccc}
			& \textbf{$\boldsymbol{E_{beam}}$} & \textbf{$\boldsymbol{G}$} & \textbf{$\boldsymbol{l}$} & \textbf{$\boldsymbol{L^*}$} & \textbf{$\boldsymbol{\beta^{*}}$} & \textbf{$\boldsymbol{\beta_{max}}$} & \textbf{$\boldsymbol{\sigma_{max}}$} \\ 
                        & (TeV) & (T/m) & (m) & (m) & (m) & (km)  & (mm)\\
			\hline Tevatron & 0.98 & 141/138 & 3.35 & 7.6 & 0.28 & 1.4 & 1.7 \\  
			LHC & 7 & 205 & 5.5 & 23 & 0.55 & 4 & 1.4 \\ 
			VLHC $pp$ & 50 & 220/190 & 20/17.5 & 36 & 1.10 & 40 & 1.6 \\  
			VLHC $p\bar{p}$ & 50 & 605/354 & 16.75/14.7 & 7.6 & 0.14 & 27 & 1.1 \\ 
		\end{tabular} 
	\end{ruledtabular}
%}
\end{table}
	
\section{Conclusions}

Large rare event cross sections make a high luminosity, 100 TeV proton-antiproton collider appealing. The high rare event rates  allow one to lower synchrotron radiation deposits into superconducting magnets and to reduce the number of events per beam crossing. A long tunnel permits simple 4.5\,T magnets and reduced stored magnetic field energy. Larger tunnels have already been successfully bored.  Parallel transverse stochastic pre-cooling is the main upgrade. Antiprotons are plentiful.

\section*{Acknowledgements}

We would like to thank  Estia Eichten, Henry Frisch, Keith Gollwitzer, Valeri Lebedev, Peter McIntyre, and Steve Mrenna for helpful information.

%%%%%%%%%%%%%%%%%%%%%%%%%%%%%%%%%%%%%%%%%%%%%%%%%%%%%%%%%%%%%%%%%%%%%%%%%%%%%%%%%%%%%%%%%%%%%%%%%%%%%%%%%%%%%%%%

\end{document}